\documentclass[aps,prb,twocolumn,showpacs,amsmath,superscriptaddress,10pt]{revtex4-1}
\usepackage{amssymb}
\usepackage{amsmath}
\usepackage{graphicx}
\usepackage{float}
\usepackage{ulem}
\usepackage{bm}
\usepackage{dcolumn}
\usepackage{mathrsfs}
\usepackage{subfigure}
\usepackage{braket}
\usepackage{graphicx}
\usepackage{braket}
\usepackage{color}
\usepackage{bbold}
\usepackage{caption}
\usepackage[colorlinks=True,citecolor=blue,linkcolor=blue,allcolors=blue]{hyperref}
\usepackage{enumitem}
\newcommand{\cosinv}{\cos^{-1}}
\usepackage{array, makecell}
\newcolumntype{C}[1]{>{\centering\arraybackslash}p{#1}}

\usepackage[skip=1ex]{caption}
\def\blue{\textcolor{black}}

\def\bea{\begin{eqnarray}}
\def\eea{\end{eqnarray}}
\def\bal{\begin{aligned}}
\def\eal{\end{aligned}}

\begin{document}
\title{Dynamical characterization of Weyl nodes in Floquet Weyl semimetal phases}

\author{Muhammad Umer}
\affiliation{Department of Physics, National University of Singapore, Singapore 117551, Republic of Singapore.}
\author{Raditya Weda Bomantara}
\email{Raditya.Bomantara@sydney.edu.au}
\affiliation{Centre for Engineered Quantum Systems, School of Physics, University of Sydney, Sydney, New South Wales 2006, Australia.}
\author{Jiangbin Gong}
\email{phygj@nus.edu.sg}
\affiliation{Department of Physics, National University of Singapore, Singapore 117551, Republic of Singapore.}

\begin{abstract}
Due to studies in nonequilibrium (periodically-driven) topological matter, it is now understood that some topological invariants used to classify equilibrium states of matter do not suffice to describe their nonequilibrium counterparts.  Indeed, in Floquet systems the additional gap arising from the periodicity of the quasienergy Brillouin zone often leads to unique topological phenomena without equilibrium analogues.  In the context of Floquet Weyl semimetal,  Weyl points may be induced at both quasienergy zero and $\pi/T$ ($T$ being the driving period) and these two types of Weyl points can be very close to each other in the momentum space. Because of their momentum-space proximity,  the chirality of each individual Weyl point may become hard to characterize in both theory and experiments, thus making it challenging to determine the system's overall topology. In this work, inspired by the construction of dynamical winding numbers in Floquet Chern insulators, we propose a dynamical invariant capable of characterizing and distinguishing between Weyl points at different quasienergy values, thus advancing one step further in the topological characterization of Floquet Weyl semimetals. To demonstrate the usefulness of such a dynamical topological invariant, we consider a variant of the periodically kicked Harper model  (the very first model in studies of Floquet topological phases) that exhibits many Weyl points, with the number of Weyl points rising unlimitedly with the strength of some system parameters.  Furthermore, we investigate the two-terminal transport signature associated with the Weyl points. Theoretical findings of this work pave the way for experimentally probing the rich topological band structures of some seemingly simple Floquet semimetal systems. 
\\
\end{abstract}

\maketitle

\section{Introduction}\label{Sec_intro}
There has been a great surge in research on topological phases of matter after the discovery of Quantum Hall effect \cite{Kliszing1980}. In addition to topological insulators \cite{Haldane1988, Kane2005, Bernevig2006, Moore2007, FuKane2007, Fu2007,  Hsieh2008, Chen2009, Roy2009, Xia2009, Zhang2009} and superconductors \cite{Hasan2010, Qi2011}, which are charaterized by gapped bulk bands, topological semimetal \cite{Vafek2014,Huang2015, Lv2015, Lv2015a, Weng2015, Xu2015, Xu2015a, Xu2015b} phases with gapless bulk bands have also been reported. The latter exhibits band touching that may occur at isolated points \cite{Wan2011,Hosur2013,Xu2015,Xu2015a}, along a line \cite{Burkov2011,Mullen2015,Bian2016,Yu2015}, or a closed loop \cite{Chen2015,Li2018}. Depending on which of these various band touching structures is featured, such topological semimetal (SM) phases can further be categorized as topological Weyl, nodal line and nodal loop semimetals, respectively. These topological semimetal phases can be characterized in terms of valence band Chern numbers (Weyl SM) \cite{Xu2011,Hosur2013} and certain winding numbers or Berry phases along a momentum space structure (Nodal line-loop SM) \cite{Burkov2011,Li2018}. 

Isolated band touching points appearing in Weyl semimetals (Weyl points) are particularly interesting due to their linear dispersion along all three quasimomenta (thus resembling relativistic particles) and their robustness against generic perturbations \cite{Wan2011}. Such Weyl points act as the equivalent of magnetic monopoles in the momentum space, whose associated magnetic charge is equal to their chirality \cite{Nielsen1981, Hosur2013}. Due to the fermion doubling theorem \cite{Nielsen1981}, Weyl nodes always appear in pairs with opposite magnetic charges. Their topological signature is further evidenced by the existence of surface states with zero dispersion along the line connecting such pairs of Weyl nodes (Fermi arcs \cite{Wan2011,Hosur2012,Potter2014}) in finite size systems. The aforementioned features of Weyl points lead to various exotic transport properties such as chiral anomaly \cite{Adler1969,Bell1969,Zyuzin2012,Liu2013a,Burkov2014}, negative magneto-resistance \cite{Nielsen1983}, and anomalous Hall effect \cite{Burkov2011a}, to name a few. For these reasons, studies of Weyl semimetal materials and how to engineer them  have remained an active research topic up to this date.

Since the last decade, the use of periodic driving has emerged as one attractive method to engineer topological materials. It leads to a variety of novel topological phases such as Floquet topological insulators \cite{Kitagawa2010,Lindner2011, DerekPRL2012, Rechtsman2013, Wang2013Exp,Rudner2013,Asboth2014,Lababidi2014, Gong2016,
Fulga2016,Zhou2018,Umer2020,Mciver2020Exp, Coldatom}, superconductors \cite{Jiang2011,Liu2013,Tong2013, RadityaPRL08, RadityaPRB08} and semimetals \cite{Bomantara2016,Bomantara2016a,Wang2016,Gong2016b,Wang2017,
Bucciantini2017,Peri2018,Zhu2020,NC2017}. In such systems, energy is no longer a conserved quantity and is replaced by a quantity termed quasienergy, which is only defined modulo the driving frequency ($\omega$). The latter feature gives rise to the formation of quasienergy Brillouin zone (BZ), where in-gap or gapless topological edge states may emerge not only around the BZ center (quasienergy zero), but also around the BZ edge (quasienergy $\omega/2$) \cite{Nathan2015}. Consequently, the definition of new dynamical invariants \cite{Rudner2013,Asboth2014,Bomantara2016, Yao2017} is often necessary to faithfully capture all the possible edge states of Floquet topological matter under open boundaries. Finally, on a more practical side, periodic driving naturally offers an extra tunable parameter which allows the realization of distinct topological phases within the same platform.

In the context of Weyl semimetals, periodic driving enables the formation of Weyl nodes and Fermi arcs at both quasienergies zero and $\omega/2$.  Interestingly, their signatures may not be uniquely captured by the Chern numbers of $2D$ slices of the system \cite{Peri2018,Zhu2020}. Indeed,  Floquet Weyl semimetal phases can exhibit a large number of band-touching points on a $2$D slice of the momentum space. With the Weyl nodes possibly appearing in pairs of opposite chirality, neither the so-called slice Chern number approach nor the slice dynamical winding number approach in Refs. \cite{Peri2018,Zhu2020} can fully capture the change in topology \blue{[See Appendix-\ref{App_Slice}]}.  Furthermore,  quantum adiabatic charge pumping in Ref. \cite{Bomantara2016} was proposed to capture the chirality of each Weyl node. However, it was observed that two Weyl nodes of the same chirality and at different quasienergy contribute oppositely towards the total charge pumped over one adiabatic cycle. As a result, such an adiabatic charge pumping scheme is generally insufficient to distinguish between such Weyl nodes or to dynamically count the number of Weyl nodes clustered together in the momentum space.

In this paper, we proposed a means for separately probing the chirality of Weyl nodes at quasienergy zero and $\omega/2$ by extending the domain of use of the dynamical winding number proposed in Ref.~\cite{Rudner2013}, which was originally proposed to characterize Floquet anomalous topological insulators. This is accomplished by evaluating such a winding number with respect to a closed surface in the three-dimensional ($3D$) BZ enclosing Weyl points under consideration. The usefulness of our proposal becomes clearer in systems exhibiting many Weyl nodes at quasienergy zero and $\omega/2$ packed very closely to one another. In such cases, a given closed surface in the $3D$ BZ may in practice enclose at least a pair of Weyl nodes with different quasienergy values, resulting in a zero net band Chern number. On the other hand, dynamical winding number calculations still yield nontrivial values which address two such Weyl points individually.  Moreover, we study two-terminal transport signatures associated with the Weyl nodes of opposite chirality.  We shall reveal that the two-terminal conductance captures the total chirality of the Weyl nodes at quasienergy zero and $\omega/2$.

The article is structured in the following way. In Sec.~\ref{Sec_Chirality_Winding}, we briefly review the literature on the Floquet theory and dynamical winding number.  This is to make this work more self consistent.  To explicitly demonstrate the correlation between dynamical winding number surrounding Weyl points and their chirality, we then consider a simple Floquet four band toy model exhibiting two Weyl nodes of different quasienergy values at the same quasimomenta.  In Sec.~\ref{Sec_KHM}, we employ the kicked Harper model, a celebrated dynamical model in the literature of quantum chaos, to further demonstrate the usefulness of dynamical winding number calculations in systems with potentially high Weyl node density. In Sec.~\ref{Sec_Conductance}, we study the two-terminal conductance associated with the Weyl nodes of opposite chirality.  Finally, we conclude our findings in Sec.~\ref{Sec_Sum}.

\section{Chirality of Weyl nodes}\label{Sec_Chirality_Winding}

\subsection{Floquet Theory: A Review}\label{Subsec_Floquet}
The Floquet theory \cite{Shirley1965, Sambe1973} is a powerful tool to study time periodic systems whose dynamics is governed by the one period unitary evolution operator, usually referred to as the Floquet operator. For a time-periodic Hamiltonian $H({\bf k},t)$ with $H({\bf k},t) = H({\bf k},t+T)$, where ${\bf k}$ is the set of system parameters (e.g., quasimomenta) and $t$ is time, the Floquet operator is denoted by $U({\bf k})$ and given as, $U({\bf k}) = \hat{\mathbb{T}}e^{-\frac{i}{\hbar}\int_{0}^{T}H({\bf k}, t)dt}$, where $\hat{\mathbb{T}}$ is time ordering operator and $T = \frac{2\pi}{\omega}$ is the time-period ($\omega$ = driving frequency) after which the Hamiltonian repeats itself. It satisfies the Floquet eigenvalue equation $\hat{\mathbb{T}}e^{-\frac{i}{\hbar}\int_{0}^{T}\hat{H}({\bf k}, t)dt}\mid\Psi_{n}({\bf k})\rangle = e^{-i\Omega_{n}({\bf k})T/\hbar}\mid\Psi_{n}({\bf k})\rangle$, where $\Omega_{n}({\bf k})$ is called quasienergy, which replaces the role of energy in such non-equilibrium systems. The quasienergy is defined modulo $w = \frac{2\pi}{T}$, which in this paper is taken $\in (\frac{-\pi}{T},\frac{\pi}{T}]$. As a consequence of this periodicity, quasienergy bands may close not only at quasienergy zero, but also at $\pi/T$. In the context of Floquet Weyl semimetals, this enables the formation of Weyl nodes at either quasienergy zero or $\pi/T$. 

\subsection{Dynamical winding number as topological invariant }\label{Subsec_Winding}

Dynamical winding number $W^\epsilon$ has been introduced in \cite{Rudner2013} to characterize the net chirality of edge states crossing a gap around quasienergy $\epsilon$ \cite{Rudner2013, Lababidi2014, Zhou2018, Umer2020}. In Floquet topological insulators, it can uniquely characterize systems with arbitrary number of co-propagating edge states \cite{Zhou2018}. Together with some additional invariants, it can further count the number of counter-propagating edge states \cite{Fulga2016, Umer2020, Lababidi2014}, thus recovering the notion of bulk-boundary correspondence in Floquet systems. The general applicability of dynamical winding number, as well as its ability to characterize a variety of Floquet topological phases with no static counterparts, has led us to think of more possibilities where it can play a significant role. As will be demonstrated in the next few sections, such an invariant can in fact be utilized to separately probe the chirality of the Weyl nodes at zero and $\pi/T$ quasienergy. To this end, we will first review the theory of dynamical winding number to develop some intuitions.

In order to calculate dynamical winding number, cyclic evolution is introduced by employing a modified time-evolution operator in momentum representation which is denoted by $\tilde{U}_{\epsilon}({\bf \Theta},t)$ and given as \cite{Rudner2013},
\bea\bal
\tilde{U}_{\epsilon}({\bf \Theta},t) =
\begin{cases}
U({\bf \Theta}, 2t) ~~&\text{if}~~ 0 \le t  < T/2\\
e^{-iH^{\epsilon}_{\rm eff}(2T - 2t)} ~~&\text{if}~~ T/2 \le t  < T \;,
\end{cases}
\label{EQ:ReturnMap}
\eal\eea
where $T = 1$ is the period of drive and $\Theta$ is the set of continuous parameters which can form a closed surface. In $2D$ systems, $\Theta = (k_{x}, k_{y})$ simply represents a set of quasi-momenta in two spatial directions, whereas in three dimensions ($3D$), $\Theta = (\theta, \phi)$ can be taken as comprising the polar and azimuthal angles that form a closed $2D$ spherical or toroidal surface in $3D$ BZ. $H^{\epsilon}_{\rm eff} = -\frac{i}{T}\log^{\epsilon}[U({\bf \Theta},T)]$ is the effective Hamiltonian, with $\epsilon$ being the branch cut of logarithm function, such that its eigenvalues $\Omega \in [\epsilon-2\pi, \epsilon]$ \cite{Rudner2013}. The operator during the second half of the drive is a return map, which sends the modified time-evolution operator to identity at the end of one period, i.e., $\tilde{U}_{\epsilon}({\bf \Theta}, t=0) = \tilde{U}_{\epsilon}({\bf \Theta}, t=T) = \mathbf{1}$. 

With the above notations, we are now ready to define the dynamical winding number $W^\epsilon$ with respect to quasienergy $\epsilon$ \cite{ Rudner2013}. By focusing in particular to $3D$ systems, it is defined as 

\bea
\bal
W^{\epsilon} &= \frac{1}{8\pi^{2}}{ \int_0^T} dt~{ \int_{\mathcal{S}}}d\theta ~d\phi \\
&\times Tr\Biggl( \tilde{U}^{-1}_{\epsilon}\partial_{t}\tilde{U}_{\epsilon}\Big[\tilde{U}^{-1}_{\epsilon}\partial_{\phi}\tilde{U}_{\epsilon}, \tilde{U}^{-1}_{\epsilon}\partial_{\theta}\tilde{U}_{\epsilon}\Big] \Biggl)\;,
\label{EQ:Winding}
\eal
\eea
where $\theta$ and $\phi$ are the polar and azimuthal angles respectively, which together parameterize a $2D$ spherical or toroidal surface $\mathcal{S}$ in the $3D$ BZ. From Eq. (\ref{EQ:ReturnMap}), we can observe that the modified Floquet operator during first half of the period depends on the driving protocol of the periodically driven system, whereas during the second half of the period, modified Floquet operator depends on the full period time-evolution operator along with the choice of branch cut of the logarithm function. Despite the seemingly complex expression of $W^{\epsilon}$, it physically counts the number of the system's full time-evolution eigenphase $\epsilon$ singularities in the effective 3D Brillouin zone spanned by $(\theta,\phi,t)\in\left(0,\pi\right]\times \left(0,2\pi\right] \times \left(0,T\right]$ space \cite{Nathan2015}. In particular, if the effective 2D surface on which Eq.~(\ref{EQ:Winding}) is computed encloses a Weyl point at quasienergy $\epsilon$, diagonalizing $U(\Theta,t)$ and plotting the phase $\tilde{\varepsilon}$ of its eigenvalues against $\theta$, $\phi$, and $t$ in the effective 3D Brillouin zone will yield a single band touching at $\tilde{\varepsilon}_0=\epsilon$. In the remainder of this paper, numerical evaluation of Eq.~(\ref{EQ:Winding}) is carried out by direct numerical integration over a 3D grid with $N_{\theta} \times N_{\phi} \times N_{t}$ discretization points. In some simple cases, we are also able to benchmark our numerics with analytical results. We remark that an alternative numerical method presented in Ref.~\cite{Hockendorf2017} may also be used to evaluate Eq.~(\ref{EQ:Winding}) without resorting to numerical integration.

\subsection{Toy model and Weyl nodes}
\label{Subsec:Toy}

In order to illustrate how the dynamical winding number defined above works in capturing the chirality of Weyl nodes, we consider a simple four band toy model. In particular, it possesses two Weyl nodes, one with quasienergy zero and the other $\pi/T$, located at the same point $(k_{x0},k_{y0},k_{z0})$ in the $3D$ BZ. The Hamiltonian of the system is defined as $H(k_{x},k_{y},k_{z}, t)$ and given as
\bea\bal
&H(k_{x}, k_{y}, k_{z}, t) = 
Jk_{x}\tau_{0}\sigma_{x} + Jk_{y}\tau_{0}\sigma_{y}\\
&~~~~~~~~~~~ + \left[-J k_{z}\tau_{+} + \pi(1 + J k_{z})\tau_{-}\right]\sigma_{z}\delta(t-nT)
~~~~~~\label{EQ:Toy_Effective}
\eal\eea
where $\tau_i,~\sigma_{i}$ are the set of Pauli matrices, $\tau_{+} = \frac{\tau_{0} + \tau_{z}}{2}$, and $\tau_{-} = \frac{\tau_{0} - \tau_{z}}{2}$ form the upper and lower diagonal matrices. The Floquet operator associated with the time periodic Hamiltonian of the system [Eq. (\ref{EQ:Toy_Effective})] for time period $t \in [0_{-},T_{-})$, where $T = \hbar = 1$ is given by
\bea\bal
U(k_{x}, k_{y}, k_{z}) = \tau_{+}\otimes U^{0} + \tau_{-}\otimes U^{\pi},
\label{EQ:Toy}
\eal\eea
where $U^{0}$ and $U^{\pi}$ can be regarded as the time evolution operators of some effective Hamiltonian possessing a single Weyl node at quasienergy zero and $\pi/T$, respectively and they are given by,
\bea\bal
U^{0}(k_{x}, k_{y}, k_{z}) &= e^{-i[ J k_{x}\sigma_{x} + J k_{y}\sigma_{y} - J k_{z}\sigma_{z}]},\\
U^{\pi}(k_{x}, k_{y}, k_{z}) &= e^{-i[\pi\sigma_{0} + J k_{x}\sigma_{x} + J k_{y}\sigma_{y} +  J k_{z}\sigma_{z}]},
\label{EQ:Toy_Part}
\eal\eea
Moreover, effective Weyl Hamiltonian $H^{0}_{\rm eff}$~[$H^{\pi}_{\rm eff}$] of these Floquet operator can be obtained by $H^{\epsilon}_{\rm eff} = -i{\rm log}[U^{\epsilon}]$. Given a Weyl Hamiltonian $H=f_x k_x\sigma_x+f_y k_y\sigma_y +f_z k_z\sigma_z$, the chirality of its associated Weyl node is given as $\chi = \mathrm{sgn}[f_{x}f_{y}f_{z}]$ \cite{Hosur2013}. In this case, the chirality of the two Weyl nodes associated with Eq.~(\ref{EQ:Toy_Part}) is then given as $\chi^{0} = -1$ and $\chi^{\pi} = +1$ for the effective Weyl Hamiltonian $H^{0}_{\rm eff}$ and $H^{\pi}_{\rm eff}$ respectively.

\begin{figure}[ht]
\centering
\includegraphics[clip, trim=0.2cm 0.5cm 0.3cm 0.3cm, width=0.96\linewidth, height=1.00\linewidth, angle=270]{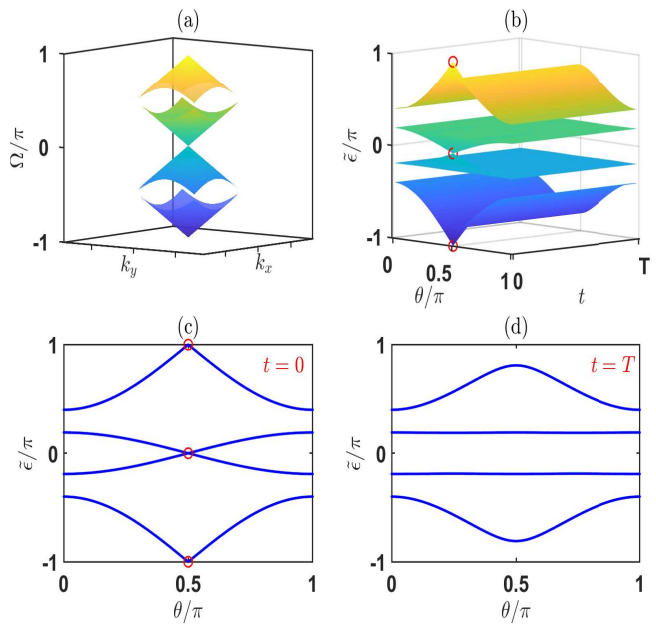}
\caption{The quasienergy spectrum of the toy model has been shown in (a) for fixed $k_{z} = 0$. ~ $(b,~c,~d)$ shows the eigen-phase band spectrum of the toy model for fixed $\phi = \pi/2$,~$(c)$~ $t = 0$ and $(d)$~ $t = T$. Weyl nodes in Floquet bands $(a)$ and phase band $(b,~c)$ crossing can be observed. Phase band crossing occur at single point in time domain which is captured by dynamical winding number.}
\label{Fig:Toy}
\end{figure}

Fig.~\ref{Fig:Toy}(a) depicts the quasienergy band structure associated with the above model. There, two Weyl nodes with quasienergy zero and $\pi/T$ are clearly observed at the same point $(0,0,0)$ in the 3D BZ. In order to directly compute the dynamical winding number on a spherical surface enclosing the Weyl nodes, we carry out coordinate transformation from Cartesian to spherical polar coordinates: $k_{x} = r\sin{\theta}\cos{\phi}, k_{y} = r\sin{\theta}\sin{\phi}$ and $k_{z} = r\cos{\theta}$, where $r$ is the radius of the sphere and it is taken to be small such that some kind of first-order approximation in our analytical treatment holds i.e; $r^2 \approx 0$.

Let $|\Psi_i\rangle$ and $\Omega_i^\epsilon \in [\epsilon-2\pi,\epsilon]$ be the $i$-th band eigenvectors and the associated quasienergy of the Floquet operator [Eq.~(\ref{EQ:Toy})]. Here $\epsilon$ is the branch cut of the logarithmic function such that the quasienergy is taken as $\Omega^{\epsilon}_{i} \in \{-2\pi+\epsilon, \epsilon\}$. We may then construct the modified Floquet operator in the spirit of Eq.~(\ref{EQ:ReturnMap}) which is given as,
\begin{flushleft}
\bea\bal
\tilde{U}_{\epsilon}({\bf \Theta},t) =
\begin{cases}
{\mathcal{T} e^{-i \int_0^{2t} dt' H(\theta,\phi, t')}} &\text{if}~ 0 \le t  < T/2\\
\sum_{i} e^{-i\Omega_{i}^{\epsilon}[2T-2t]}\mid\Psi_{i}\rangle\langle\Psi_{i}\mid  &\text{if}~ T/2 \le t  < T \;.
\end{cases}
~~~\label{EQ:6}
\eal\eea
\end{flushleft}
which is unitary such that $\tilde{U}_{\epsilon}\tilde{U}^{-1}_{\epsilon} = \tilde{U}^{-1}_{\epsilon} \tilde{U}_{\epsilon} = \mathbb{1}$. The dynamical winding number can be determined by dividing the time integral into two parts from $t\in[0,T/2)$ and $t\in[T/2,T)$. The Eq. (\ref{EQ:Winding}) during the time interval $t\in[0,T/2)$ will be given as,
\bea\bal
W^{\epsilon}(t_{0\rightarrow\frac{T}{2}}) &= \frac{1}{8\pi^{2}}{ \int_{0}^{T/2}} dt~{ \int_{\mathcal{S}}}d\theta ~d\phi \\
&\times Tr\Biggl( \tilde{U}^{-1}_{\epsilon}\partial_{t}\tilde{U}_{\epsilon}\Big[\tilde{U}^{-1}_{\epsilon}\partial_{\phi}\tilde{U}_{\epsilon}, \tilde{U}^{-1}_{\epsilon}\partial_{\theta}\tilde{U}_{\epsilon}\Big] \Biggl)\
\eal\eea
which leads to {$W^{\epsilon}(t_{0\rightarrow\frac{T}{2}}) = 0$} as $\cos(r)\approx{1}$ and $\sin(r)\approx{r}$ under our ``first-order'' approximation. That is, for cases with sufficiently small $r$, it becomes clear that the dynamical winding number is only contributed by $\tilde{U}_\epsilon$ during the time interval $t\in [T/2,T)$. The modified Floquet during this interval is given as,
\bea \bal
\tilde{U}_{\epsilon}(\theta, \phi, t_{T/2\rightarrow{T}}) = \tau_{+}\otimes \tilde{U}^{1,2}_{\epsilon} + \tau_{-}\otimes \tilde{U}^{3,4}_{\epsilon},
\label{EQ:4band_Floquet}
\eal \eea 
where
\newline
\begin{widetext}
\bea\bal
   \tilde{U}^{1, 2}_{\epsilon}&=
  \left( {\begin{array}{cc}
   e^{-i\Omega_{1}^{\epsilon}[2T-2t]}\sin^{2}(\frac{\theta}{2}) + e^{-i\Omega_{2}^{\epsilon}[2T-2t]}\cos^{2}(\frac{\theta}{2}) &~~ \frac{e^{-i\phi}\sin(\theta)}{2} (e^{-i\Omega_{1}^{\epsilon}[2T-2t]} - e^{-i\Omega_{2}^{\epsilon}[2T-2t]} ) \\
   \frac{e^{i\phi}\sin(\theta)}{2} (e^{-i\Omega_{1}^{\epsilon}[2T-2t]} - e^{-i\Omega_{2}^{\epsilon}[2T-2t]} ) &~~  e^{-i\Omega_{1}^{\epsilon}[2T-2t]}\cos^{2}(\frac{\theta}{2}) + e^{-i\Omega_{2}^{\epsilon}[2T-2t]}\sin^{2}(\frac{\theta}{2}) \\
  \end{array} } \right), \\
\\
  \tilde{U}^{3, 4}_{\epsilon}&=
  \left( {\begin{array}{cc}
   e^{-i\Omega_{3}^{\epsilon}[2T-2t]}\sin^{2}(\frac{\theta}{2}) + e^{-i\Omega_{4}^{\epsilon}[2T-2t]}\cos^{2}(\frac{\theta}{2}) &~~ -\frac{e^{-i\phi}\sin(\theta)}{2} (e^{-i\Omega_{3}^{\epsilon}[2T-2t]} - e^{-i\Omega_{4}^{\epsilon}[2T-2t]} ) \\
   -\frac{e^{i\phi}\sin(\theta)}{2} (e^{-i\Omega_{3}^{\epsilon}[2T-2t]} - e^{-i\Omega_{4}^{\epsilon}[2T-2t]} ) &~~  e^{-i\Omega_{3}^{\epsilon}[2T-2t]}\cos^{2}(\frac{\theta}{2}) + e^{-i\Omega_{4}^{\epsilon}[2T-2t]}\sin^{2}(\frac{\theta}{2}) \\
  \end{array} } \right),
\eal\eea
\end{widetext}

The dynamical winding number during time interval $t\in[T/2,T)$ is {then} given as, 
\begin{flushleft}
\bea \bal
W^{\epsilon}(t_{{T/2}\rightarrow{T}}) &= \frac{1}{8\pi^{2}}{\int_{0}^{\pi}}d\theta ~{\int_{0}^{2\pi}}d\phi{\int_{T/2}^{T}}dt\\
& \times Tr\Biggl( \tilde{U}^{-1}_{\epsilon}\partial_{t}\tilde{U}_{\epsilon}\Big[\tilde{U}^{-1}_{\epsilon}\partial_{\phi}\tilde{U}_{\epsilon}, \tilde{U}^{-1}_{\epsilon}\partial_{\theta}\tilde{U}_{\epsilon}\Big] \Biggl)\;,~~~~~\\
& = -\frac{1}{2\pi}\big[-\Omega_{1}^{\epsilon} + \Omega_{2}^{\epsilon} - \Omega_{3}^{\epsilon} + \Omega_{4}^{\epsilon} \\ &+ \sin{(\Omega_{1}^{\epsilon}-\Omega_{2}^{\epsilon})} + \sin{(\Omega_{3}^{\epsilon}-\Omega_{4}^{\epsilon})}\big].
~~~~~\label{EQ:Toy_result}
\eal \eea
\end{flushleft}
where $\epsilon$ is the choice of the branch cut of logarithmic function and it is taken as either $0$ or $\pi/T$. Moreover,  $\Omega_{i}^{\epsilon} (i\in [1,2,3,4])$ is the quasienergy of the $i^{th}$ band and depends on the choice of the branch cut $\epsilon$ of the logarithmic function. For $\epsilon = 0$, the quasienergy is taken $\Omega\in[-2\pi,0)$ and we will have $ \Omega_{1}^{0} = -2\pi + \tan^{-1}{(Jr)}, \Omega_{2}^{0} = -\tan^{-1}{(Jr)},  \Omega_{3}^{0} = -\pi - \tan^{-1}{(Jr)}$ and $\Omega_{4}^{0} = -\pi + \tan^{-1}{(Jr)}$ which then produces $\chi^{0} = W^{0} = -1$ from Eq.~ (\ref{EQ:Toy_result}). Similarly, for the quasienergy gap or the branch cut $\epsilon = \pi$, the quasienergy is taken in the period of $-\pi$ to $\pi$ and the quasienergy of the bands are given as $ \Omega_{1}^{\pi} = \tan^{-1}{(Jr)}, \Omega_{2}^{\pi} = -\tan^{-1}{(Jr)},  \Omega_{3}^{\pi} = \pi - \tan^{-1}{(Jr)}$ and $\Omega_{4}^{\pi} = -\pi + \tan^{-1}{(Jr)}$ which then produces $\chi^{\pi} = W^{\pi} = +1$ from Eq. (\ref{EQ:Toy_result}). These results are in full agreement with the chirality determined for the effective Hamiltonian from Eq. (\ref{EQ:Toy_Part}) at zero and $\pi$ quasienergy gaps.

In Fig. \ref{Fig:Toy}(b), we plot the eigenphase spectrum of the system's full time-evolution operator surrounding its Weyl points,
\begin{equation}
U(\theta,\phi,t) = \mathcal{T} e^{-\mathrm{i} \int_0^{t} dt'\;H(\theta,\phi,t')} \;,
\end{equation} 
at $\phi = \pi/2$ slice. We observe that band crossing at both $\tilde{\varepsilon}=0,\pi$, for $\theta = \pi/2$ which lies on the equator of the sphere for an arbitrary value of $\phi$. This band crossing circle can be continuously moved to the pole forming a single band crossing point but it can not be removed without changing the topology of the system at quasienergy $\epsilon$. This confirms the physical interpretation of $W^\epsilon$ \cite{Nathan2015} elucidated in Sec.~\ref{Subsec_Winding}. In Fig. \ref{Fig:Toy} $(c,~d)$, we plot the phase bands for fixed time of $t = 0$ and $t = T$ where Fig. \ref{Fig:Toy} $(b,~d)$ shows phase band opening for $t > 0$.

The above analysis illustrates the mechanism in which dynamical winding number captures the chirality of Weyl nodes at zero and $\pi/T$ quasienergy located at a shared single point in 3D BZ. Due to the system's simplicity, the calculated dynamical winding number can be directly compared to the Weyl points' chirality obtained from inspecting the Hamiltonian Eq.~(\ref{EQ:Toy_Effective}). In other more complicated Floquet Weyl semimetals, inspecting $\chi^0$ and $\chi^\pi$ directly from their definition with respect to the effective Weyl Hamiltonian may no longer be analytically feasible. Moreover, in systems capable of hosting as many Weyl nodes as wish, such as that considered in the following section, these Weyl nodes may necessarily be packed too close to one another. Consequently, isolating a single Weyl point and evaluating its chirality via some Chern number related response measurement are not practically feasible. These represent scenarios in which our proposed dynamical winding characterization becomes an extremely useful tool to probe the systems' various Weyl points.

\section{Kicked Harper Model}\label{Sec_KHM}
In this section, we investigate  a variant of the so-called kicked Harper model as a rich model of Floquet topological matter \cite{Leboeuf1990KHM,Wang2013KHM,Derek2014KHM,Bomantara2016}. Note that the kicked Harper model was a seminal dynamical model in the literature of quantum chaos and it is actually the first model ever used to examine topological phase transitions in Floquet quasienergy bands \cite{Leboeuf1990KHM}.  The Hamiltonian in the lattice basis can be written as  
\bea \bal
&\hat{H}_{\text{KHM}} = \sum_{n = 1}^{N-1}\sum_{j} V\cos(2\pi\beta_{2}n + \alpha_{z}) \mid n\rangle\langle n\mid\delta(t-jT) \\
& + \sum_{n = 1}^{N-1}\bigl[J + \lambda\cos(2\pi\beta_{1}n + \alpha_{y})\bigl]\mid n+1\rangle\langle n\mid +H.c. \label{EQ:HKHM}
\eal \eea
where $n$ represent the lattice site index while $N$ is the total number of lattice sites in the system. $J$ and $\lambda$ are the hopping parameters and {$V$ is the kicking field strength}. $t$ is the time while $T$ is the time period of the drive. $\beta_{1}$ and $\beta_{2}$ are the two parameters which determine the periodicity of the lattice system in two artificial dimensions of $\alpha_{y}$ and $\alpha_{z}$ respectively. By fixing $\beta_{1} = \beta_{2} = 1/2$, we obtain a two band system in which the $\alpha_{y}$ and $\alpha_{z}$ represent the quasi-momenta in two artificial dimensions respectively. The Hamiltonian in the momentum representation is then given as, 
\bea\bal
&H_{\text{KHM}}(k_{x}, \alpha_{y}, \alpha_{z}, t) = 2J\cos(k_{x})\sigma_{x}\\ ~~~~~& + 2\lambda\sin(k_{x})\cos(\alpha_{y})\sigma_{y} + V\cos(\alpha_{z})\sigma_{z}\delta(t-jT)
~~~\label{EQ:KHM_HAM}
\eal\eea
where $\sigma_j$ are the Pauli matrices in the sublattice degree of freedom and $k_{x}$ is the momentum along the physical dimension. 

We can easily write the system's Floquet operator as (by considering the time interval $t \in \{0_{-}, T_{-}\}$) 
\bea\bal
U_{\text{KHM}}&(k_{x}, \alpha_{y}, \alpha_{z}) \\ = &e^{-i[2J\cos(k_{x})\sigma_{x} + 2\lambda\sin(k_{x})\cos(\alpha_{y})\sigma_{y}]}e^{-iV\cos(\alpha_{z})\sigma_{z}}
~~~\label{EQ:KHM_Floquet}
\eal\eea
where we have again fixed $\hbar = T = 1$. It is worth mentioning that the detailed analysis of the above model has been studied in \cite{Bomantara2016}, with Weyl and line nodes, as well as nodal loops explicitly identified at certain parameter values. 

In this paper, we focus on the regime for which Weyl nodes exist and calculate the dynamical winding number and Floquet band Chern number surrounding these points. Here and in the remainder of this paper, we refer to the Chern number associated to the lower quasienergy band defined in the quasienergy Brillouin zone $\left(-\pi/T,\pi/T\right]$. It is also to be emphasized that such a band Chern number is well-defined since Weyl nodes are enclosed and {\it not} within the surface on which such a quantity is computed. To identify the regime for which Weyl nodes exist, we first note that $U_{\rm KHM}$ can be easily diagonalized, which yields two quasienergies $\Omega_{\pm} =  \pm \cos^{-1}\bigl[\cos[f_{1}]~\cos[f_{2}]\bigl]$, where $f_{1} = V\cos(\alpha_{z})$ and $f_{2} = \sqrt{4J^{2}\cos^{2}(k_{x}) + 4\lambda^{2}\cos^{2}(\alpha_{y})\sin^{2}(k_{x})}$. It thus follows that band touching can only occur at either zero or $\pi/T$ quasienergy for $f_2=0$, $f_{1} = 2\ell\pi$ or $f_1=(2\ell+1)\pi$ respectively, where $\ell\in \mathbb{Z}$. The pinning of the band touching at zero or $\pi/T$ quasienergy can be understood from the emergent chiral symmetry at $f_1=\ell \pi$. In this case, the second exponential of Eq.~(\ref{EQ:KHM_Floquet}) reduces to the number $(-1)^\ell$, and the system's effective Hamiltonian can be written as
\begin{eqnarray}
H_{\rm KHM, eff}(\ell) &=& (-1)^\ell \left(2J\cos(k_x)\sigma_x\right. \nonumber \\
&& \left. +2\lambda \sin(k_x)\cos(\alpha_y)\sigma_y\right) \;,
\end{eqnarray}
which satisfies $\sigma_z H_{\rm KHM, eff}(\ell) \sigma_z = -H_{\rm KHM, eff}(\ell)$.

Reference~\cite{Bomantara2016} further found that, following such band touching events, a new set of Weyl nodes at quasienergy $2\ell \pi/T \;\mathrm{mod} \; 2\pi/T$ emerges at $(k_{x_{0}}, \alpha_{y_{0}}, \alpha_{z_{0}}) = (\pm\pi/2, \pm\pi/2, \pm\cos^{-1}[\frac{2\ell\pi}{V}])$. In particular, such a model can host as many Weyl points as we wish by tuning the parameter $V$.
\begin{flushleft}
    \begin{table}[tp]
\caption{System parameters are taken as $J = \lambda = 1$, $V = 16$ in our analysis. We consider closed $2D$ surface enclosing various number of Weyl nodes and determine their chirality in the form of dynamical winding number $(W^{0},~W^{\pi})$ while $(\mathcal{C})$ represent the Chern number. Chirality of the Weyl nodes $(\chi^{0}, \chi^{\pi})$ are directly derived from analytical solutions of the effective Hamiltonian associated with each individual Weyl node enclosed. We have denoted each case by $\Delta_{i}$ and defined $\mu_{n} = \cos^{-1}(\frac{n\pi}{V})$. $2D$ closed surface is taken in the form of a torus $(\Delta_{1}-\Delta_{6})$ such that $\delta_{x} = [R + r\sin(\theta)]\sin(\phi)$, $\delta_{y} = r\cos(\theta)$ and $\delta_{z} = [R + r\sin(\theta)]\cos(\phi)$ where $R [r]$ is the radius from the center of circle [tube] of the torus. Moreover, we consider spherical geometry of the surface for $(\Delta_{7}-\Delta_{8})$ where we consider $R = 0$ and $\theta \in [0,\pi)$. $k_x$, $\alpha_y$, and $\alpha_z$ here refer to the center of the small 2D tori or spheres used in our calculations.}
    \setcellgapes{8pt}
    \makegapedcells
\centering
\begin{tabular}{||C{4.0mm}||C{28.0mm}|C{13.0mm}|C{12.0mm}|C{14.0mm}|C{5.0mm}||}
\hline\hline
$\Delta_{i}$ & ${\tiny (k_{x},\alpha_{y},\alpha_{z})}$ & $(R, r)$ & $(\chi^{0}, \chi^{\pi})$ & $(W^{0},W^{\pi})$ & $\mathcal{C}$ \\ \hline \hline
$\Delta_{1}$ & $(\frac{\pi}{2},\frac{\pi}{2},\pm[\mu_{2}+R])$ & $(\frac{\mu_{4}}{13}, \frac{\mu_{4}}{17})$ & $(\mp{1}, 0)$ & $(\mp{1}, 0)$ & $\mp{1}$ \\ \hline
$\Delta_{2}$ & $(\frac{\pi}{2},\frac{\pi}{2},\pm[\mu_{3}+R])$ & $(\frac{\mu_{4}}{13}, \frac{\mu_{4}}{17})$ & $(0, \mp{1})$ & $(0, \mp{1})$ & $\pm{1}$\\ \hline
$\Delta_{3}$ & $(\frac{\pi}{2},\frac{\pi}{2},\pm\frac{\pi}{2})$ & $(\mu_{5}, \frac{\mu_{5}}{2})$ & $(0, \mp{2})$ & $(0, \mp{2})$ & $\pm{2}$\\ \hline
$\Delta_{4}$ & $(\frac{\pi}{2},\frac{\pi}{2},\pm\frac{\pi+2\mu_{5}}{2})$ & $(\mu_{5}, \frac{\mu_{5}}{2})$ & $(\mp{2}, 0)$ & $(\mp{2}, 0)$ & $\mp{2}$\\ \hline
$\Delta_{5}$ & $(\frac{\pi}{2},\frac{\pi}{2},\pm\frac{\pi-\mu_{5}}{2})$ & $(\frac{\mu_{4}}{5}, \frac{\mu_{5}}{2})$ & $(\mp{1}, \mp{1})$ & $(\mp{1}, \mp{1})$ & 0 \\ \hline
$\Delta_{6}$ & $(\frac{\pi}{2},\frac{\pi}{2},\pm\frac{\pi-2\mu_{5}}{2})$ & $(\frac{\mu_{2}}{4}, \mu_{5})$& $(\mp{2}, \mp{2})$ & $(\mp{2}, \mp{2})$ & 0 \\ \hline
$\Delta_{7}$ & $(\frac{\pi}{2},\frac{\pi}{2},\frac{\pm\pi}{2})$ & $(0,\frac{\mu_{2}}{4})$ & $(\mp{1},\mp{2})$ & $(\mp{1},\mp{2})$ & $\pm{1}$ \\ \hline
$\Delta_{8}$ & $(\frac{\pi}{2},\frac{\pi}{2},\pm\frac{\pi+2\mu_{5}}{2})$ & $(0,\frac{\mu_{2}}{4})$ & $(\mp{2},\mp{1})$ & $(\mp{2},\mp{1})$ & $\mp{1}$ \\ \hline \hline
\end{tabular}
\label{Table:KHM}
    \end{table}
\end{flushleft}

Let us now take $(2\ell+1) \pi< V <(2\ell+2)\pi$. The system then hosts $\ell+1$ quartets of Weyl points with zero energy at $(k_{x_{0}}, \alpha_{y_{0}}, \alpha_{z_{0}}) = (\pm\pi/2, \pm\pi/2, \pm\cos^{-1}[\frac{2q\pi}{V}])$ and $\ell+1$ quartets of Weyl points with $\pi/T$ quasienergy $(k_{x_{0}}, \alpha_{y_{0}}, \alpha_{z_{0}}) = (\pm \pi/2, \pm\pi/2, \pm\cos^{-1}[\frac{(2q+1)\pi}{V}])$, where $q=1,2,\cdots \ell+1$. We may further write the effective Weyl Hamiltonian around these Weyl points. For example, by expanding $U_{\rm KHM}$ at $(\pi/2+\delta_x, \pi/2+\delta_y, \pm\cos^{-1}[\frac{2q\pi}{V}]+\delta_z)$ and $(\pi/2+\delta_x, \pi/2+\delta_y, \pm\cos^{-1}[\frac{(2q+1)\pi}{V}]+\delta_z)$, we obtain the effective Hamiltonians 
\bea \bal
H^{0,q}_{\rm eff} &= 2q\pi\sigma_{0} - 2J\delta_{x}\sigma_{x} - 2\lambda\delta_{y}\sigma_{y} \mp \zeta_{0}\delta_{z}\sigma_{z}, \\
H^{\pi,q}_{\rm eff} &= (2q+1)\pi\sigma_{0} - 2J\delta_{x}\sigma_{x} - 2\lambda\delta_{y}\sigma_{y} \mp \zeta_{1}\delta_{z}\sigma_{z}, \label{EQ:effH}
\eal \eea
where $\zeta_{0} = \sqrt{1-\frac{4q^{2}\pi^{2}}{V^2}}$ and $\zeta_{1} = \sqrt{1-\frac{(2q+1)^{2}\pi^{2}}{V^2}}$. The chirality of these Weyl nodes can again be deduced from the effective Hamiltonian \cite{Hosur2013} and are given as $\chi^{0,\pi} = \mp{1}$ at zero and $\pi/T$ quasienergy. In Ref.~\cite{Bomantara2016}, it has been shown through quantum adiabatic pumping that when multiple Weyl nodes with quasienergy zero are enclosed in a closed surface, the total charge pumped during the adiabatic cycle captures their net chirality. On the other hand, if some enclosed Weyl nodes are of quasienergy $\pi/T$, then the total charge pumped may no longer correlate with the Weyl points' net chirality. In the following, we verify that the dynamical winding number always yields the correct net chirality in both cases.

\begin{figure*}
\centering
\includegraphics[width=0.7\linewidth,
height=1.0\linewidth,angle=270]{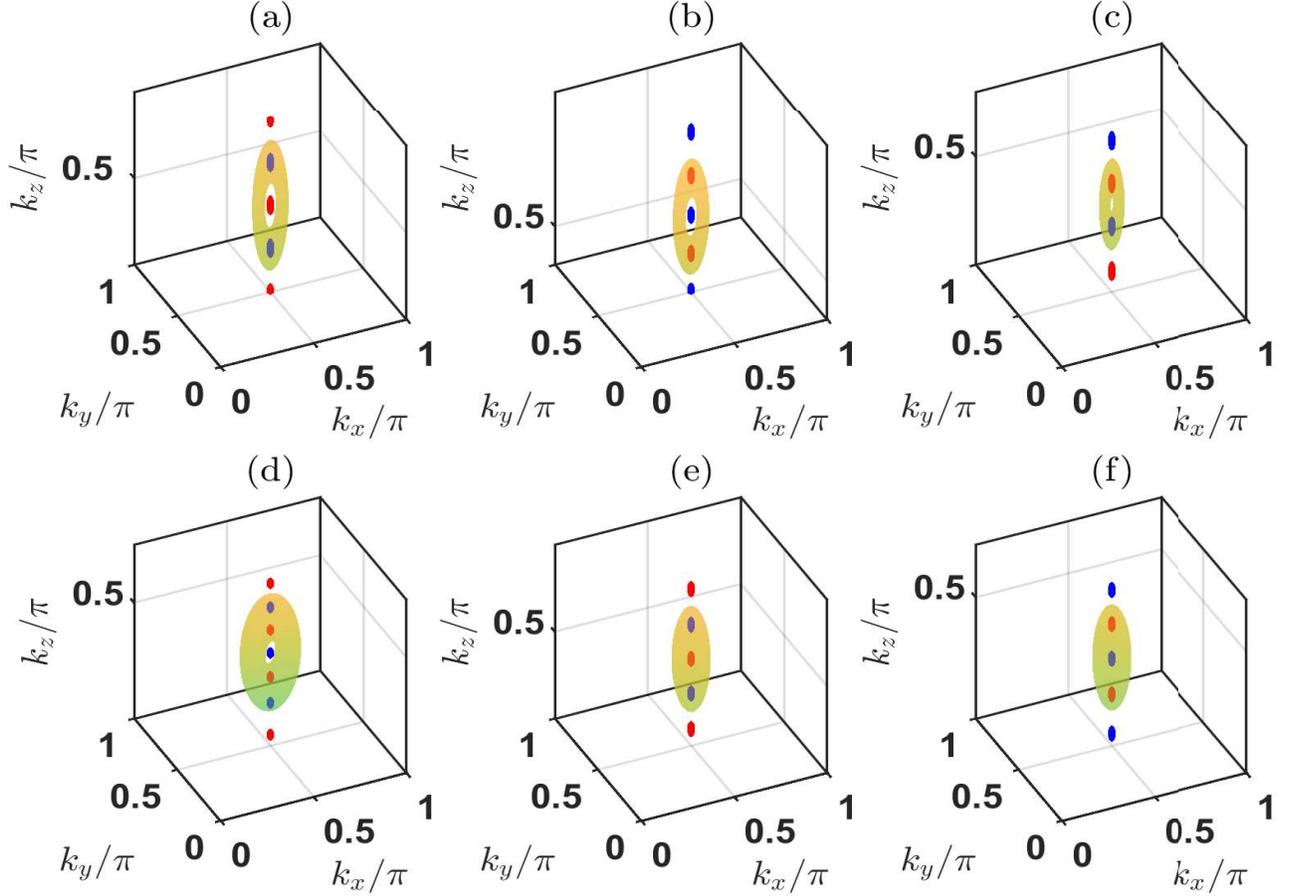}
\caption{Weyl nodes at zero (Red) and $\pi$ (Blue) quasienergy in the Brillouin zone are shown. We consider a torus (a-d) and spherical (e-f) geometry. The $2D$ torus surface is defined such that $\delta_{x} = [R + r\sin(\theta)]\sin(\phi), \delta_{y} = r\cos(\theta)$, and $\delta_{z} = [R + r\sin(\theta)]\cos(\phi)$ where $\theta$ and $\phi$ are polar and azimuthal angle $\theta, \phi \in [0,2\pi)$. For spherical geometry , we consider $R = 0$ and $\theta \in [0,\pi)$.}
\label{Fig:KHM_Torus}
\end{figure*}

We summarize our results in Table~\ref{Table:KHM} and present the analytical calculations of the dynamical winding number and Chern number in Appendix~\ref{App_WN} and Appendix~\ref{App_Chern} respectively. The dynamical winding number and Chern number are determined over a closed surface enclosing the Weyl node(s). We have considered the torus geometry which is parametrized such that $\delta_{x} =  [R + r\sin(\theta)]\sin(\phi), \delta_{y} =  r\cos(\theta)$ and $\delta_{z} =  [R + r\sin(\theta)]\cos(\phi)$, where $R~[r]$ is the radius from the center of circle [tube] of the torus and $R > r$. The torus is spanned by polar ($\theta$) and azimuthal ($\phi$) angle which are the continuous parameters and $\theta,~\phi \in [0, 2\pi)$. We have labelled various cases by $\Delta_{i}$ in Table \ref{Table:KHM} which we will discuss in detail. Let us first focus on the points where surface encloses a single Weyl node at zero or $\pi/T$ quasienergy which corresponds to the case $\Delta_{1}$ and $\Delta_{2}$ respectively in Table~\ref{Table:KHM}. There, while the dynamical winding number correctly captures the chirality of each Weyl node, the Chern number instead predicts the opposite chirality of the Weyl node at quasienergy $\pi/T$. In Appendix~\ref{App_Chern}, we highlight the origin of the minus one factor relating the Chern number and the chirality of the Weyl point at quasienergy $\pi/T$.

Secondly, we turn our attention to the situation where the surface encloses more than one Weyl nodes at a given quasienergy which is shown in Fig. \ref{Fig:KHM_Torus} (a,b). Two Weyl nodes at quasienergy $\pi/T$~[0] are shown in Fig. \ref{Fig:KHM_Torus} (a~[b]) which correspond to the case $\Delta_{3}~[\Delta_{4}]$ in Table \ref{Table:KHM}. In this case, both the dynamical winding number and Chern number yield the expected net chiralities when the two Weyl nodes are of quasienergy zero. On the other hand, if the two Weyl nodes are of quasienergy $\pi/T$, the Chern number results in the wrong sign, whereas the dynamical winding number continues to faithfully produce the correct net chirality.

Next, we turn our attention to the point labelled as $\Delta_{5}$ in Table~\ref{Table:KHM}, which corresponds to a surface enclosing two Weyl points with different quasienergy Fig. \ref{Fig:KHM_Torus}(c), but of the same chirality. In this case, the dynamical winding number correctly captures the net chirality of both Weyl nodes, whereas the Chern number instead gives zero. Similarly, the point labelled as $\Delta_{6}$ correspond to a surface enclosing two Weyl nodes at zero~$[\pi/T]$, which have the same chirality. While dynamical winding number determines the net chirality of Weyl nodes at zero and $\pi/T$ quasienergy, the Chern number $\mathcal{C} = W^{0} - W^{\pi}$ \cite{Rudner2013} itself {has} no information about the chirality which can be observed from the results presented in Table \ref{Table:KHM}. 

\begin{figure*}
\centering
\includegraphics[width=0.80\linewidth, height=1.00\linewidth, angle=270]{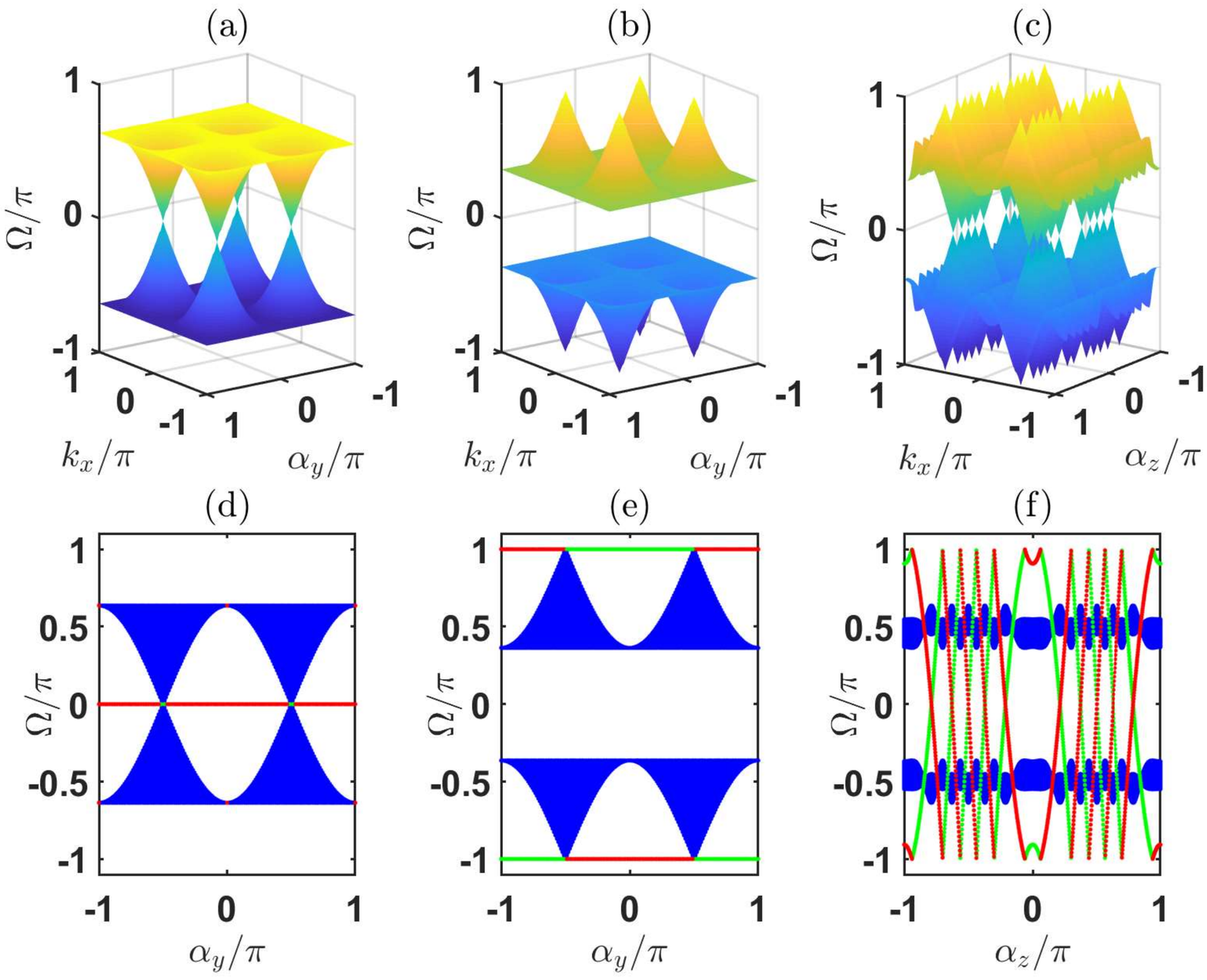}
\caption{A slice of the system's quasienergy spectrum at parameter values $J = \lambda = 1$ and $V = 16$ and (a,d) $\alpha_{z_{0}} = \pi/2$ (b,e) $\alpha_{z_{0}} = \cos^{-1}(\pi/V)$, (c) $\alpha_{y_{0}} = \pi/2$, and (f) $\alpha_{y_{0}} = \pi/4$. Panels (a,b,c) and (d,e,f) are obtained under periodic (open) boundary conditions respectively. Red (Green) color represent the states localized at the left (right) edge of the open lattice in $x-$direction.}
\label{Fig:KHM}
\end{figure*}

Finally, we consider a spherical surface such that $R = 0$ and $\theta \in [0,\pi)$ which encloses an odd number of Weyl nodes such that there is an imbalance between number of Weyl nodes at zero and $\pi/T$ quasienergy, see $\Delta_{7}-\Delta_{8}$ in Table~\ref{Table:KHM}. The sphere in Fig. \ref{Fig:KHM_Torus}(e) depicts the situation where the surface encloses one~[two] Weyl node at quasienergy zero~[$\pi/T$] and refers to point $\Delta_{7}$ in the Table \ref{Table:KHM}. Similarly, a surface encloses one~[two] Weyl node at $\pi/T$~[zero] quasienergy has been shown in Fig. \ref{Fig:KHM_Torus}(f) which correspond to point $\Delta_{8}$ in Table \ref{Table:KHM}. The dynamical winding number captures the net chirality while the Chern number once again provide the difference of Weyl nodes at zero and $\pi/T$ quasienergy. The above analysis emphasizes on the dynamical winding number characterization of the Weyl nodes in Floquet Weyl semimetals. 

Before ending this section, we verify the presence of Fermi arcs in the system when OBC are applied in one direction. In particular, we focus on a parameter regime for which many Weyl {points at quasienergy} zero and $\pi/T$ coexist,  which are hence very close to each other in 3D Brillouin zone. Our results are summarized in Fig.~\ref{Fig:KHM}. By plotting the quasienergy spectrum at two different $\alpha_{z_0}=\pi/2$ and $\alpha_{z_0}=\cos^{-1}\left(\frac{\pi}{V}\right)$ values, Fermi arcs at quasienergy zero and $\pi/T$ can be observed in panels (d) and (e) respectively. The Fermi arcs connect the two band touching points through both the BZ center and edge (e.g., degenerate edge states are present both at $\alpha_y=0$ and $\alpha_y=\pi$). This is possible due to the fact that each band touching point observed in Fig.~\ref{Fig:KHM}(d) or (e) corresponds to the projection of two Weyl points at $k_x=\pm \pi/2$ in Fig.~\ref{Fig:KHM}(a) or (b) respectively to the system's surfaces, where each pair of Weyl points thus contributes to each of the two Fermi arcs that together span the whole $\alpha_y$ BZ. Moreover, since the system hosts Weyl points that appear in quartets due to the presence of time-reversal symmetry, the Chern number on any fixed $\alpha_z$ plane is zero. This is further evidenced in Fig.~\ref{Fig:KHM}(f) that the system's quasienergy spectrum at a fixed $\alpha_{y}=\pi/4$ plane yields counter-propagating chiral edge states at both quasienergy zero and $\pi/T$. These counter-propagating chiral edge states can be captured through two-terminal conductance \cite{Umer2020} which signals that the Weyl nodes of opposite chirality might have the same transport response which is studied in the next section.

The above results further demonstrate the application of dynamical winding number in categorizing the Floquet Weyl semimetal phases. In particular, the cases $\Delta_{3}-\Delta_{8}$ in Table~\ref{Table:KHM} represent the scenario for which dynamical winding number calculation is truly necessary for probing the presence of coexisting Weyl nodes at quasienergy zero and $\pi/T$. Strictly speaking, in two-band systems, it is impossible for two Weyl nodes at zero and $\pi/T$ quasienergy to coincide at the same quasimomenta. However, certain systems, such as that considered in this section, are capable of hosting a large number of Weyl nodes. Consequently, due to the limited size of the 3D BZ, these Weyl nodes may necessarily be very close to one another [which can be observed in Fig. \ref{Fig:KHM}(c)]. In this case, considering a small enough closed surface that encloses only a single Weyl point will be difficult to achieve in practice. We expect that this is the scenario for which the proposed dynamical winding number calculation will be most useful.

\section{Two-Terminal Conductance and total chirality of Weyl nodes}\label{Sec_Conductance}
In the previous section, we have studied that the dynamical winding number efficiently captures the net chirality of Weyl nodes enclosed by a surface. It is evident that the dynamical winding number will not capture the total chirality of the Weyl nodes, i.e., the total number of Weyl points. Furthermore, though the dynamical winding number determines the net chirality of the Weyl nodes, it cannot distinguish between a single Weyl node and three Weyl nodes, two of which having opposite chirality. Such subtleties require the information regarding the total number of Weyl nodes for a thorough characterization of topological entities, i.e., Weyl nodes in this case. Indeed, this may be understood as another interesting aspect of nonequilibrium topological matter. 

\begin{figure}[H]
\centering
\includegraphics[clip, trim=8.0cm 0.5cm 7.0cm 0.5cm, width=0.25\linewidth, height=1\linewidth, angle=270]{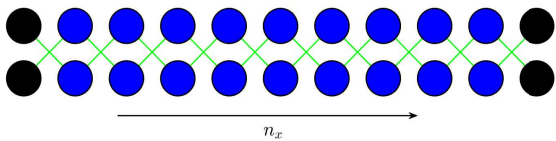}
\caption{Schematic of a one dimensional lattice in the $x$-direction with $N_{x}$ number of unit cells for our two-terminal conductance analysis. Each unit cell consists of two orbital degrees of freedom and green bonds represent inter-cell coupling. We applied absorbing boundary conditions in the form of zero dimensional point contacts (black) at $n_{x} = 1$ and $n_{x} = N_{x}$.}
\label{Fig:Lattice}
\end{figure}

In this section, we attempt to capture the total chirality of the Weyl nodes through conductance signatures in two-terminal transport. For this purpose, we use the Floquet scattering matrix approach \cite{Fulga2016}, which can be applied in a straightforward manner.  We consider a finite lattice, with $q$ orbital degrees of freedom, in the physical axis for some fixed $\alpha_{y}$ and $\alpha_{z}$ as tunable parameters. Moreover, we apply point-like absorbing terminals at the ends of the lattice as shown in Fig. \ref{Fig:Lattice}. The projector on to the absorbing leads is chosen as,
\bea \bal 
P = \begin{cases}
1 ~~~ &if~~ n_{x} \in \{1,N_{x}\}\;,\\
0 ~~~ &\text{otherwise}\;,
\end{cases}
\eal \eea
where $n_{x}$ is the lattice site index. The projector acts stroboscopically.  That is to say that, the absorbing terminals only act at the beginning and end of each period. This condition can be achieved provided that sufficient control over the leads-system interaction is possible. For example,  in a previous experiment, this control was accomplished by the use of tunable external gates that can effectively render the links connecting the leads and the system, hence the interaction between them, insulating or conducting \cite{Gallagher2014}. As further discussed below,  in a more natural setting 
where the system and the leads are not subject to periodic driving, the conductance results from the scattering matrix approach still have a clear physical meaning.

The unitary scattering matrix of dimension $2q\times{2q}$ is then defined as $S^{\epsilon}$ and given by,
\bea\bal 
S^{\epsilon} = P\left[\mathbb{1} - e^{i\epsilon}\hat{U}(1-P^{T}P)\right]^{-1}e^{i\epsilon}\hat{U}P^{T} \;, \label{EQ:scatter}
\eal\eea
where $T$ denotes the matrix transpose, $\epsilon$ is the quasienergy gap $\epsilon \in \{0,\pi\}$ and $\hat{U}$ being the Floquet operator under the boundary conditions defined above. The resulting ${2q}\times{2q}$ scattering matrix becomes the following:
\bea
\bal
S^{\epsilon} = \left(\begin{array}{cc} r & t\\ t^{*} & r^{*} \end{array}\right) \;,
\eal
\eea
where $^{*}$ corresponds to the complex conjugation, $r$ and $t$ are the $q\times{q}$ blocks of reflection and transmission amplitudes respectively. The two-terminal conductance is then given as a function of quasienergy as $G^{\epsilon} = \text{Trace}(tt^{*})$, where $\epsilon$ is taken in either zero or $\pi$ gap. In actual scattering experiments, an incoming state cannot be prepared at a given quasienergy $\epsilon$ value. Instead, it usually possesses a given energy. In this case, a single conductance measurement experiment may not yield the expected quantized results below. However, Refs.~\cite{Sumrule,Yap2017, Yap2018} have demonstrated that such a quantization can be recovered by repeating the experiments for different incoming energy values and then applying the so-called {\it Floquet sum rule}~\cite{Sumrule}. 

As we change periodic boundary conditions to open boundary conditions, the Weyl nodes at $(k_{x}, \alpha_{y}, \alpha_{z}) = (\pm\pi/2, \alpha_{y}, \alpha_{z})$ at quasienergy $\epsilon$ project themselves at the surface of the system. First of all, we choose $(\alpha_{y}, \alpha_{z}) = (\pm\pi/2, \pm\cosinv[\frac{2q\pi}{V}])$ such that two Weyl nodes of opposite chirality exist at zero quasienergy $\epsilon = 0$. The two-terminal conductance is found to yield $(G^{0},G^{\pi}) = (2,0)$, which captures the total chirality of the Weyl nodes at zero quasienergy whereas zero value of $G^{\pi}$ indicates that there is no Weyl node at $\pi$ quasienergy. Secondly, we consider the tunable parameter of artificial dimension such that Weyl nodes occur at $\pi$ quasienergy for $(\alpha_{y}, \alpha_{z}) = (\pm\pi/2, \pm\cosinv[\frac{(2q+1)\pi}{V}])$. The two-terminal conductance of the system is found to be $(G^{0},G^{\pi}) = (0,~2)$ where it predicts the total chirality of the Weyl nodes at $\pi$ quasienergy. 
\begin{figure}
\centering
\includegraphics[width=1.0\linewidth, height=1.0\linewidth, angle=270]{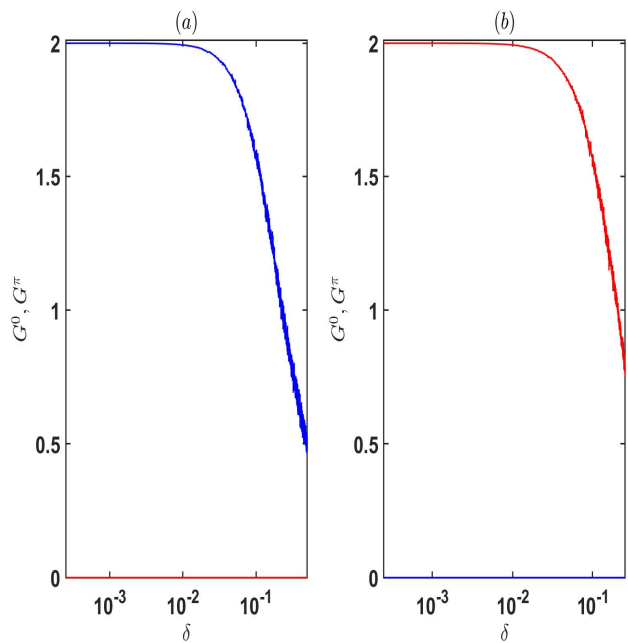}
\caption{Two-terminal conductances $(G^{0}, G^{\pi})$ at zero (blue) and $\pi/T$ (red) quasienergy of a one dimensional lattice in the $x$-direction with $N_{x} = 101$ number of unit cells. We consider $J = \lambda = 1, V = 16$,  $\alpha_{y} = \pm\pi/2$, (a) $\alpha_{z} = \pm\cosinv[\frac{2\ell\pi}{V}]$, and (b) $\alpha_{z} = \pm\cosinv[\frac{(2\ell + 1)\pi}{V}]$. All points are obtained after averaging over 1000 disorder realizations.}
\label{Fig:Disorder}
\end{figure}

To demonstrate the robustness of the calculated two-terminal conductance, we repeat the above analysis with respect to a disordered lattice. More precisely, we modify the system's Hamiltonian of Eq.~(\ref{EQ:HKHM}) to implement disorder in all parameter values, resulting in  
\begin{small}
\bea \bal
&\hat{H}_{\text{KHM}} = \sum_{n = 1}^{N-1}\sum_{j} V(1+\delta^{1}_{n})\cos(2\pi\beta_{2}n + \alpha_{z}) \mid n\rangle\langle n\mid\delta(t-jT) \\
& + \sum_{n = 1}^{N-1}\bigl[J(1+\delta^{2}_{n}) + \lambda(1 +\delta^{3}_{n}) \cos(2\pi\beta_{1}n + \alpha_{y})\bigl]\mid n+1\rangle\langle n\mid +H.c.
\eal \eea
\end{small}
\noindent where $\delta^{i}_{n}$ is derived from a uniform distribution such that $\delta^{i}_{n} \in \{-\delta, \delta\}$ where $\delta$ is the strength of the disorder. We present our result in Fig.~\ref{Fig:Disorder}, where each point is averaged over 1000 disorder realizations. A plateau around $G^0,G^\pi=2$ is clearly observed at small disorder strengths, thus confirming the robustness of the two terminal conductances $G^0$ and $G^\pi$ that represent the Weyl node's chirality at quasienergy zero and $\pi/T$ respectively. At moderate disorder strengths, we observe that the two-terminal conductances deviate polynomially from their expected quantized value. Such breakdown in conductance quantization can be understood from the fact that moderate and strong disorder may lead to hybridization of the two bands around quasienergy zero and/or $\pi/T$. In this case, the associated two-terminal conductance loses its topological nature.

The above results demonstrate the capability of $G^{\epsilon}$ to capture the total chirality of the Weyl nodes at quasienergy $\epsilon$ even in the presence of sufficiently small system imperfections (disorder). Indeed, both chiral \cite{Zhou2018} and counter-propagating \cite{Umer2020} surface states associated with the Weyl nodes contribute positively to the two-terminal conductance, thus providing the necessary information regarding the total chirality of the Weyl nodes. To summarize this section, we have shown that the Weyl nodes' total chirality at quasienergy $\epsilon$ can be probed by $G^\epsilon$, whereas their net chirality can instead be captured by evaluating the appropriate dynamical winding number.

\section{concluding remarks}\label{Sec_Sum}
In this paper, we have proposed the use of dynamical winding number to characterize the Weyl points in Floquet Weyl semimetal phases. Using a simple four band toy model, we demonstrate how dynamical winding number can separately address Weyl points at quasienergy zero and $\pi/T$ when they are located at the same point in the $3D$ BZ. To further compare the usefulness of dynamical winding number with that of Chern number in the context of probing Weyl points, we analyse a variant of the seminal kicked Harper model as a Floquet Weyl semimetal. Our investigation reveals that the dynamical winding number over a closed $2D$ surface (which has been chosen to be either of spherical or toroid shape) always correctly determines the net chirality of all the Weyl points enclosed (regardless of their quasienergy). By contrast, when such a surface encloses multiple Weyl points of different quasienergy values, the Chern number does not reflect the net chirality of the multiple Weyl points under investigation. Moreover, we have studied the two-terminal transport signature associated with the Weyl points of opposite chirality. It is found that the two-terminal conductance captures the total magnitude of the chirality of Weyl nodes at zero and $\pi/T$ quasienergy. 

\begin{center}
	\begin{figure}
		\includegraphics[scale=0.5]{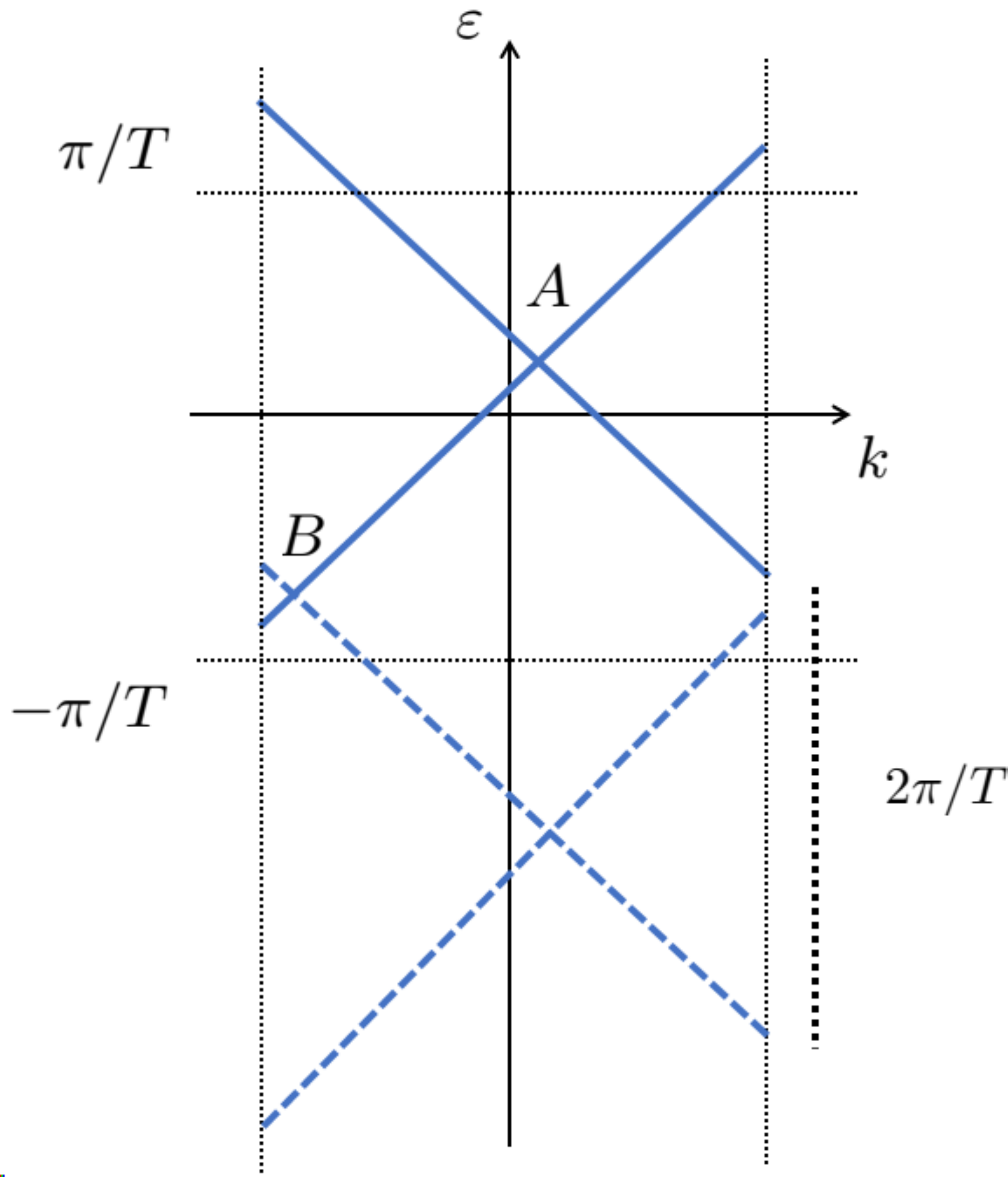}
		\caption{Two distinct species of Weyl points (marked as $A$ and $B$) may arise in general time-periodic Weyl semimetal systems. Solid lines depict the dispersion within the main BZ of $(-\pi/T, \pi/T]$, with $A$ being the band touching point within the main BZ.  Dashed lines represent the equivalent of solid lines by shifting $2\pi/T$, with $B$ a result of the crossing between different Floquet side bands.   This feature is unique in Floquet systems because the quasi-energy is only defined up to $2\pi/T$. }
		\label{Fig:Wgen}
	\end{figure}
\end{center}

It should also be emphasized that while the system we considered above only admits Weyl points at either quasienergy zero or $\pi/T$, the dynamical invariant and two terminal conductance studies we proposed also provide similar advantages over the usual Chern number analysis in the general time-periodic setting with Weyl points occurring at any quasienergy. In particular, even in this general setting, two distinct species of Weyl points may still arise due to the periodicity of the quasienergy Brillouin zone, i.e., the Weyl point labelled A (B) in Fig.~\ref{Fig:Wgen} forms when the center band touches the other band from below (above). That is, Weyl node of type A is a result of band crossing inside the same Floquet sideband, whereas Weyl point of type B emerges from the crossing between different Floquet sidebands.  In the presence of particle-hole symmetry, Weyl point A (B) is thus pinned at quasienergy zero ($\pi/T$).

Note that while the role of the two types of Weyl points can in principle be exchanged by a global quasienergy shift, the wrong choice of quasienergy band can lead to an additional negative sign in the chirality (Chern number) of Weyl node. Namely, suppose that evaluating the Chern number along the blue coloured band around the Weyl point A gives exactly its chirality. In this case, evaluating the Chern number along the same blue coloured band but around the Weyl point B will introduce an extra -1 factor to the expected chirality. A similar situation occurs if one chooses to evaluate the Chern number along the red coloured band instead, in which case it captures the exact chirality of Weyl point B, but now introduces an extra -1 factor to the chirality of Weyl point A. Theoretically, Chern number calculation is sufficient to characterize the topology of any Weyl points (A and B) in Floquet Weyl semimetals, i.e., by isolating a very small $2$D surface enclosing a Weyl point of interest and computing the Chern number with respect to the appropriate band. In practice, however, the execution of this procedure may not be straightforward for the following two main reasons. First and foremost, given a system with multiple Weyl points that are very close to one another, it is not easy to construct a sufficiently small $2$D surface that only encloses a single Weyl point. Second, we might not know in practice which choice of band leads to the Chern number giving the correct chirality without the introduction of -1 factor. On the other hand, the dynamical invariants and two terminal conductance above can address such problems. In this more general setting, such quantities can still be defined by choosing appropriate $\epsilon$ in Eqs.~(\ref{EQ:Winding}) and (\ref{EQ:scatter}). More importantly, it is not necessary to fine tune $\epsilon$ at a specific value where a Weyl point under consideration resides. In general, given a 2D surface enclosing a Weyl point, choosing $\epsilon$ anywhere inside a relevant quasienergy gap of a Floquet operator defined on such a surface is expected to work well.

Finally, we will briefly comment on a relevant past study \cite{Sun2018} of Floquet Weyl semimetal involving Weyl points of species A and B in Fig.~\ref{Fig:Wgen}. There, the authors introduced a winding number $\nu_3$ that captures the collective property of both species of Weyl points in terms of their net handedness. Such a winding number is fundamentally different from the dynamical winding number defined in the present study, which {\it separately} captures the handedness/chirality of each Weyl point. The potential advantage of the dynamical winding number over $\nu_3$ of Ref.~\cite{Sun2018} becomes appreciable in systems with many Weyl points such as that studied in the present paper. In such systems,  $\nu_3=1$ may correspond to, e.g., the setting of one type A Weyl point and one type B Weyl point of the same chirality, or a rather complication setting with two type A Weyl points of the same chirality plus two type B Weyl points of opposite chirality. By contrast, our dynamical winding number $W^{\epsilon}$ is capable of distinguishing between these different scenarios.

As a possible future study, it would be interesting to look into the dynamical characterization of other Floquet Weyl semimetal phases. Some of these possibilities are the Floquet type-II Weyl semimetal and Floquet multi Weyl semimetal phases where dynamical winding number is expected to capture the higher monopole charges and chiralities associated with each Weyl node. Secondly, it would also be of much interest to study whether the dynamical winding number can characterize phases of higher-order Weyl semimetal phases.

\vspace{0.5cm}
\acknowledgments
{It is a pleasure to acknowledge helpful discussions with Longwen Zhou and Linhu Li.} R.W.B is supported by the Australian Research Council Centre of Excellence for Engineered Quantum Systems (EQUS, CE170100009). J. Gong acknowledges support from Singapore National Research Foundation Grant No. NRF- NRFI2017-04 (WBS No. R-144-000-378-281).

\onecolumngrid
\appendix
\vspace{0.5cm}
\section{Slice topological behaviour of Floquet Weyl semimetals}\label{App_Slice}
{In this section, we discuss the behaviour of the slice Chern and winding numbers. This allows us to demonstrate the advantage of the method presented in Sec.~\ref{Subsec_Winding} (winding number calculation over a closed surface) over slice Chern number or slice winding number methods. First, we note that the slice Chern number procedure, which refers to the process of scanning the Chern number (with respect to two quasimomenta ($k_{x},k_{y}$)) over all values of the third quasimomentum $k_{z}$ may fail to capture Weyl points topology of a system possessing more than two Weyl points with opposite chirality at the same quasimomentum $k_{z_0}$. Extending this procedure to the winding number setting, i.e., the slice winding number approach considered in Ref. ~\cite{Zhu2020}, one may now address Weyl points at quasienergy zero and $\pi/T$ separately, but the main problem remains. Namely, while the slice winding number procedure is capable of characterizing a system possessing a Weyl point at quasienergy zero and another Weyl point at quasienergy $\pi/T$, both of which are located at the same $k_{z_0}$, it is still unable to characterize a system with two Weyl points (at the same quasienergy) of opposite chirality at the same $k_{z_0}$.}

\begin{figure}[!htb]
\centering
\includegraphics[width=0.3\linewidth, height=0.45\linewidth, angle=270]{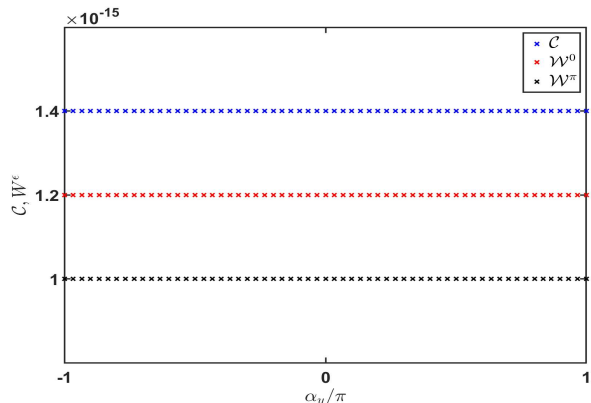}
\caption{\blue{The numerical results for the slice Chern number and slice dynamical winding number in the extended KHM for system parameter values $J = \lambda = 1,~V = 16$. It can be observed that there is no jump at the transition point which are given as $\alpha_{y} = \pm \pi/2$. Chern number and dynamical winding numbers show trivial behaviour and these negligible values are due to some numerical error.}}
\label{Fig:Slice_CW}
\end{figure}

\blue{In the extended KHM model which is considered in Sec.~\ref{Sec_KHM}, many Weyl points at quasienergy zero and $\pi/T$ may emerge along any of the three quasimomentum directions. Therefore, both slice Chern number and slice winding number calculations mentioned above will typically fail to capture the system\textquotesingle{s} topology. To support this statement, we have explicitly calculated both quantities, by scanning over all quasimomentum $\alpha_{y}$ values [See Fig.~\ref{Fig:KHM}(d, e)], in extended KHM system and summarize our results in Fig.~\ref{Fig:Slice_CW}. It can be observed that both the slice Chern number and the dynamical winding number do not exhibit any jump while sweeping $\alpha_{y}$ from $-\pi$ to $\pi$. It can be explained as follows. The Weyl nodes at a given $\alpha_{y} = \alpha_{y_{0}}$ appear in pairs with opposite chirality for $k_{x} = -\pi/2$ and $k_{x} = \pi/2$. That is, if Weyl nodes of positive chirality are appearing at $(k_{x}, \alpha_{y}, \alpha_{z}) = (-\pi/2, \alpha_{y_0}, \alpha_{z_0})$ then there is another Weyl node at $(k_{x}, \alpha_{y}, \alpha_{z}) = (\pi/2, \alpha_{y_0}, \alpha_{z_0})$ which has negative chirality. Thus, the net chirality of the Weyl nodes at a given quasienergy for $\alpha_{y} = \alpha_{y_0}$ is zero, which is why the slice Chern number and the slice dynamical winding number do not change their values across $\alpha_{y} = \alpha_{y_0}$. Furthermore, if one considers quasimomentum $k_{x}~ \left[\alpha_{z}\right]$ as the sweeping direction, similar observations can be made because for fixed $k_{x_0}~\left[\alpha_{z_0}\right]$ pairs of Weyl nodes with opposite chirality appear in the system.}

\blue{For completeness, we also present here two different examples where the slice Chern number or slice dynamical winding number does capture the topology. A static Weyl semimetal obtained from stacking Chern insulators represents a system in which slice Chern number does capture its Weyl point topology. With the Hamiltonian given by $H({\bf k}) = \sin(k_{x})\sigma_{x} + \sin(k_{y})\sigma_{y} + \left[M + \cos(k_{x}) + \cos(k_{y}) + \cos(k_{z})\right]\sigma_{z}$ for $M = - 2.5$, the Weyl nodes will appear at $(k_{x}, k_{y}, k_{z}) = (0, 0, \pm \pi/3)$. In this example, the slice Chern number $\mathcal{C}(k_{z})$ will take a jump from zero to $\pm 1$ at $k_{z} = -\pi/3$ and from $\pm 1$ to zero at $k_{z} = \pi/3$, hence capturing the change in the topology of the underlying $2$D effective system due to the presence of Weyl points. A more non-trivial example in which slice winding number is able to capture the system topology is presented in Ref.~\cite{Zhu2020}. It can be observed in Fig.~3c(ii) of Ref.~\cite{Zhu2020}, that Weyl nodes appear for $(k_{x}, k_{y}, k_{z}) = (0, 0, \pm{0.48\pi})$ at quasienergy $\epsilon$ and $(k_{x}, k_{y}, k_{z}) = (\pi, \pi, \pm{0.67\pi})$ at quasienergy $\epsilon + \pi/T$. Here at a given $k_{z}$ value, a single Weyl node appears at either $\epsilon$ or $\epsilon + \pi/T$ quasienergy hence the slice dynamical winding number does capture the change in the topology effectively. Moreover, one can observe in Fig.~3c(ii) of Ref.~\cite{Zhu2020} that even slice Chern number is capturing the change in the topology of the underlying $2$D effective system.}

\section{Dynamical Winding number calculation}\label{App_WN}
In this section, we carry out the simplest analytical calculation of dynamical winding number by considering a variant of the kicked Harper model presented in section \ref{Sec_KHM}. We consider the Weyl node in the three dimensional Brillouin zone $(k_{x_{0}}, \alpha_{y_{0}}, \alpha_{z_{0}}) = (\pi/2, \pi/2, \pi/2)$ such that $(k_{x}, \alpha_{y}, \alpha_{z}) = (k_{x_{0}} + \delta_{x}, \alpha_{y_{0}} + \delta_{y}, \alpha_{z_{0}} + \delta_{z})$ where $\delta_{x}, \delta_{y}$ and $\delta_{z}$ are the deviations from the Weyl point in three spatial directions. Moreover, we consider $\delta_{x} = r\sin(\theta)\cos(\phi), ~\delta_{y} = r\sin(\theta)\sin(\phi)$ and $\delta_{z} = r\cos(\theta)$ which form a closed $2D$ surface around the Weyl point in the form of a sphere, where $r$ is taken small such that we may use some kind of first-order approximation to obtain the dynamical winding number with convenience. 

By expanding the time-dependent Hamiltonian around this point for $2J = 2\lambda = V = f$ (for simplicity), we obtain
\bea\bal
H(\theta, \phi, t) &= -fr\sin(\theta)\cos(\phi)\sigma_{x} - f{r}\sin(\theta)\sin(\phi)\sigma_{y}\\ 
&- fr\cos(\theta)\sigma_{z}\delta(t-jT),
\label{EQ:App_Ham}
\eal\eea
The Floquet operator is then given as,
\bea\bal
U(\theta,\phi) = e^{i[\sigma_{0} + fr\sin(\theta)\cos(\phi)\sigma_{x} + f{r}\sin(\theta)\sin(\phi)\sigma_{y} + fr\cos(\theta)\sigma_{z}]}
\eal\eea
where $U(\theta, \phi)$ is the Floquet operator with $\Omega_{i}$ and $\mid\Psi_{i}\rangle$ being the quasienergy and eigenvectors which are given as, 
\[ \mid\Psi_{1}\rangle = \left( \begin{array}{c}
-e^{-i\phi}\sin(\frac{\theta}{2})  \\
\cos(\frac{\theta}{2}) 
\end{array} \right)~~~~~~~
\mid\Psi_{2}\rangle = \left( \begin{array}{cc}
e^{-i\phi}\cos(\frac{\theta}{2})  \\
\sin(\frac{\theta}{2}) 
\end{array} \right)
\]
\newline
with quasienergy $\Omega^{0}_{1} = -2\pi + \tan^{-1}(fr)$, $\Omega^{0}_{2} = - \tan^{-1}(fr)$ which is defined in the range $\Omega \in [-2\pi ,0)$ with branch cut of logarithmic function $\epsilon = 0$. Similarly for branch cut $\epsilon = \pi$, the quasienergy are given as $\Omega^{\pi}_{1} = + \tan^{-1}(fr)$, $\Omega^{\pi}_{2} = - \tan^{-1}(fr)$ which is defined in the range $\Omega \in [-\pi ,\pi)$. 
The modified Floquet operator during the time interval $t\in [0,T/2)$ is then given as,
\bea\bal
\tilde{U}^{\epsilon}(\theta,\phi,2t) = e^{i[fr\sin(\theta)\cos(\phi)\sigma_{x} + fr\sin(\theta)\sin(\phi)\sigma_{y}]2t}e^{ifr\cos(\theta)\sigma_{z}}
\eal\eea
which leads to $W^{\epsilon}(t_{0\rightarrow{T/2}}) = 0$, up to first order in $r$. 

The modified Floquet operator during the time interval $t\in[T/2,T)$ is given as $\tilde{U}(\theta,\phi,t_{T/2\rightarrow{T}}) = \sum_{i=1}^{2}e^{-i\Omega^{\epsilon}_{i}[2T-2t]}\mid\Psi_{i}\rangle\langle\Psi_{i}\mid$ which results in, 
\bea\bal\label{EQ:App_modified4}
\tilde{U}(\theta,\phi, t_{\frac{T}{2}\rightarrow{T}}) = \left( {\begin{array}{cc}
   e^{-i\Omega^{\epsilon}_{1}[2T-2t]}\sin^{2}(\frac{\theta}{2}) + e^{-i\Omega^{\epsilon}_{2}[2T-2t]}\cos^{2}(\frac{\theta}{2}) &~~ \frac{e^{-i\phi}\sin(\theta)}{2} (-e^{-i\Omega^{\epsilon}_{1}[2T-2t]} + e^{-i\Omega^{\epsilon}_{2}[2T-2t]} ) \\
   \frac{e^{i\phi}\sin(\theta)}{2} (-e^{-i\Omega^{\epsilon}_{1}[2T-2t]} + e^{-i\Omega^{\epsilon}_{2}[2T-2t]} ) &~~  e^{-i\Omega^{\epsilon}_{1}[2T-2t]}\cos^{2}(\frac{\theta}{2}) + e^{-i\Omega^{\epsilon}_{2}[2T-2t]}\sin^{2}(\frac{\theta}{2}) \\
  \end{array} } \right),
\eal\eea
which leads to the dynamical winding number $W^{\epsilon}(t_{\frac{T}{2}\rightarrow{T}})$ given as,
\bea\bal
W^{\epsilon}(t_{\frac{T}{2}\rightarrow{T}}) =  \frac{\Omega^{\epsilon}_{1} - \Omega^{\epsilon}_{2} - \sin(\Omega^{\epsilon}_{1} - \Omega^{\epsilon}_{2})}{2\pi},
\eal\eea
where $\Omega^{0}_{1} = -2\pi + \tan^{-1}(fr)$ and $\Omega^{0}_{2} = - \tan^{-1}(fr)$. This results in $W^{0}= -1$ for $r$ being a small number and captures the chirality of the Weyl node. On the other-hand, for the branch cut $\epsilon = \pi$, the quasienergy is found to be $\Omega^{\pi}_{1} = + \tan^{-1}(fr)$ and $\Omega^{\pi}_{2} = - \tan^{-1}(fr)$, yielding $W^{\pi} = 0$ and hence that the $\pi$ quasienergy gap does not have a Weyl node. 

In summary, we have shown that up to some kind of first-order approximation in treating a small 2D closed surface, the dynamical winding number can be directly calculated and it is found to capture the chirality of the chosen Weyl nodes here at zero quasienergy. Similar calculation can be carried out for the Weyl node at quasienergy $\pi/T$.

\section{The origin of the extra sign between the Chern number and chirality of the Weyl point at quasienergy $\pi/T$}
\label{App_Chern}
The main idea of this section is to note that the system's effective Hamiltonian can be obtained via two distinct approaches. On the one hand, if we expand the time-dependent Hamiltonian around a Weyl point before computing its Floquet operator, we end up with the effective Hamiltonian presented in Eq.~(\ref{EQ:effH}), which allows the explicit determination of the Weyl point's chirality. On the other hand, Chern number is calculated with respect to eigenstates of Floquet operator, thus suggesting that one should first evaluate the system's Floquet operator before expanding it around the Weyl point to obtain the associated effective Hamiltonian. Specifically, by noting that the system's full Floquet operator takes the form 
\bea\bal 
U_{\rm KHM}(k_{x},\alpha_{y},\alpha_{z}) = d_{0}\sigma_{0} - i[d_{x}\sigma_{x} + d_{y}\sigma_{y} + d_{z}\sigma_{z}],
\eal\eea
where
\bea \bal 
d_{0} &= \cos(\alpha)\cos(V\cos(\alpha_{z})),\\
d_{x} &= \frac{2\sin(\alpha)}{\alpha}\big[J\cos(k_{x})\cos(V\cos(\alpha_{z})) + \lambda\cos(\alpha_{y})\sin(k_{x})\sin(V\cos(\alpha_{z}))\big],\\
d_{y} &= \frac{2\sin(\alpha)}{\alpha}\big[\lambda\cos(\alpha_{y})\sin(k_{x})\cos(V\cos(\alpha_{z})) - J\cos(k_{x})\sin(V\cos(\alpha_{z}))\big],\\
d_{z} &= \cos(\alpha)\sin(V\cos(\alpha_{z})),
\eal\eea
expanding it around the Weyl nodes at zero and $\pi/T$ quasienergy located at $(k_{x_{0}}, \alpha_{y_{0}}, \alpha_{z_{0}}) = (\frac{\pi}{2}, \frac{\pi}{2}, \pm\cosinv[\frac{2q\pi}{V}])$ and $((\frac{\pi}{2}, \frac{\pi}{2}, \pm\cosinv[\frac{(2q+1)\pi}{V}])$ respectively yields
\bea\bal
U^{0,q}(\delta_{x}, \delta_{y} \delta_{z}) &= \cos(2q\pi)\big[\sigma_{0} - i[-2J\delta_{x}\sigma{x} -2\lambda\delta_{y}\sigma{y} \mp \zeta_{0}\delta_{z}\sigma_{z}]\big],\\
U^{\pi,q}(\delta_{x}, \delta_{y} \delta_{z}) &= \cos((2q+1)\pi)\big[\sigma_{0} - i[-2J\delta_{x}\sigma{x} -2\lambda\delta_{y}\sigma{y} \mp \zeta_{1}\delta_{z}\sigma_{z}]\big] \;.
\eal\eea
These then result in the effective Hamiltonians, 
\bea \bal 
\tilde{H}^{0,q}_{\rm eff} &= 2q\pi\sigma_{0} - 2J\delta_{x}\sigma_{x} - 2\lambda\delta_{y}\sigma_{y} \mp \zeta_{0}\delta_{z}\sigma_{z}, \\
\tilde{H}^{\pi,q}_{\rm eff} &= (2q+1)\pi\sigma_{0} + 2J\delta_{x}\sigma_{x} + 2\lambda\delta_{y}\sigma_{y} \pm \zeta_{1}\delta_{z}\sigma_{z}.
\eal \eea
By comparing these results with Eq.~(\ref{EQ:effH}), we observe that a relative negative sign in the coefficients of the Pauli matrices appears between $\tilde{H}^{0,q}_{\rm eff}$ and $H^{0,q}_{\rm eff}$. This in turn explains the relative sign between the Chern number and the actual chirality of the Weyl point at quasienergy $\pi/T$.

To summarize,  for a Weyl point at zero quasienergy, the Chern number is given 
by the lowest quasienergy band with respect to zero and gives the correct chirality. However, for a Weyl point with quasienergy $\pi/T$, the correct chirality is given by the Chern number of the above band 
(or effectively, the associated band should be right below the $\pi/T$ gap).
\twocolumngrid
\bibliographystyle{apsrev4-2}

\begin{thebibliography}{99}


\bibitem{Kliszing1980}
K. v. Klitzing, G. Dorda and M. Pepper,
\prl~{\bf  45}, 494 (1980).

\bibitem{Haldane1988}
F. D. M. Haldane,
\prl~{\bf 61}, 2015 (1988).

\bibitem{Kane2005}
C. L. Kane and E. J. Mele,
\prl~{\bf 95}, 226801 (2005).

\bibitem{Bernevig2006}
B. A. Bernevig, T. L. Hughes and S.-C. Zhang,
Science {\bf 314}, 1757 (2006).

\bibitem{FuKane2007}
L. Fu and C. L. Kane,
\prb~{\bf 76}, 045302 (2007).

\bibitem{Fu2007}
L. Fu, C. L. Kane and E. L. Mele,
\prl~{\bf 98}, 106803 (2007).

\bibitem{Moore2007}
J. E. Moore and L. Balents,
\prb~{\bf 75}, 121306(R) (2007).

\bibitem{Hsieh2008}
D. Hsieh, D. Qian, L. Wray, Y. Xia, Y. S. Hor, R. Cava and M. Z. Hasan, 
Nature {\bf 452}, 970 (2008).

\bibitem{Xia2009}
Y. Xia, D. Qian, D. Hsieh, L. Wray, A. Pal, H. Lin, A. Bansil, D. Grauer, Y. S. Hor, R. J. Cava and M. Z. Hasan, 
Nat. Phys. {\bf 5}, 398 (2009).


\bibitem{Zhang2009}
H. Zhang, C.-X. Liu, X.-L. Qi, X. Dai, Z. Fang and S.-C. Zhang, 
Nat. Phys. {\bf 5}, 438 (2009).

\bibitem{Roy2009}
R. Roy,
\prb~{\bf 79}, 195322 (2009).


\bibitem{Chen2009}
Y. L. Chen, J. G. Analytis, J.-H. Chu, Z. K. Liu, S.-K. Mo, X. L. Qi, H. J. Zhang, D. H. Lu, X. Dai, Z. Fang, S. C. Zhang, I. R. Fisher, Z. Hussain, Z.-X. Shen,
Science {\bf 325}, 178 (2009).

\bibitem{Hasan2010}
M. Z. Hasan and C. L. Kane,
Rev. Mod. Phys. {\bf 82}, 3045 (2010).

\bibitem{Qi2011}
X.-L. Qi and S.-C. Zhang,
Rev. Mod. Phys. {\bf 83}, 1057 (2011).

\bibitem{Vafek2014}
O. Vafek and A. Vishwanath,
Annu. Rev. Condens. Matter Phys. ~{\bf 5}, 83
(2014).

\bibitem{Huang2015}
S.-M. Huang, S.-Y. Xu, I. Belopolski, C.-C. Lee, G. Chang, B. Wang, N. Alidoust, G. Bian, M. Neupane, C. Zhang, S. Jia, A. Bansil, H. Lin and M. Z. Hasan,
Nat. Commun. {\bf 6}, 7373 (2015).


\bibitem{Lv2015}
B. O. Lv, H. M. Weng, B. B. Fu, X. P. Wang, H. Miao, J. Ma, P. Richard, X. C. Huang, L. X. Zhao, G. F. Chen, Z. Fang, X. Dai, T. Qian, and H. Ding,
Phys. Rev. X ~{\bf 5}, 031013 (2015).

\bibitem{Lv2015a}
B. Q. Lv, N. Xu, H. M. Weng, J. Z. Ma, P. Richard, X. C. Huang, L. X. Zhao, G. F. Chen, C. E. Matt, F. Bisti, V. N. Strocov, J. Mesot, Z. Fang, X. Dai, T. Qian, M. Shi and H. Ding,
Nat. Phys. {\bf 11}, 724 (2015).

\bibitem{Weng2015}
H. Weng, C. Fang, Z. Fang, B. A. Bernevig and X. Dai,
Phys. Rev. X ~{\bf 5}, 011029 (2015).

\bibitem{Xu2015}
S.-Y. Xu, I. Belopolski, D. S. Sanchez, C. Zhang, G. Chang, C. Guo, G. Bian, Z. Yuan, H. Lu, T.-R. Chang, P. P. Shibayev, M. L. Prokopovych, N. Alidoust, H. Zheng, C.-C. Lee, S.-M. Huang, R. Sankar, F. Chou, C.-H. Hsu, H.-T. Jeng, A. Bansil, T. Neupert, V. N. Strocov, H. Lin, S. Jia and M. Z. Hasan,
Science Advances, {\bf 1}(10), e1501092 (2015).

\bibitem{Xu2015a}
S.-Y. Xu, I. Belopolski, N. Alidoust, M. Neupane, G. Bian, C. Zhang, R. Sankar, G. Chang, Z. Yuan, C.-C. Lee, S.-M. Huang, H. Zheng, J. Ma, D. S. Sanchez, B. Wang, A. Bansil, F. Chou, P. P. Shibayev, H. Lin, S. Jia, and M. Z. Hasan,
Science, ~{\bf 349}, 613 (2015).

\bibitem{Xu2015b}
S.-Y. Xu, N. Alidoust, I. Belopolski, Z. Yuan, G. Bian, T.-R. Chang, H. Zheng, V. N. Strocov, D. S. Sanchez, G. Chang, C. Zhang, D. Mou, Y. Wu, L. Huang, C.-C. Lee, S.-M. Huang, B. Wang, A. Bansil, H.-T. Jeng, T. Neupert, A. Kaminski, H. Lin, S. Jia, and M. Z. Hasan,
Nat. Phys. ~{\bf 11}, 748 (2015).

\bibitem{Wan2011}
X. Wan, A. M. Turner, A. Vishwanath and S. Y. Savrasov,
\prb ~{\bf 83}, 205101 (2011).

\bibitem{Hosur2013}
P. Hosur and X. Qi,
Comptes Rendus Physique, ~{\bf 14}, 857 (2013).

\bibitem{Burkov2011}
A. A. Burkov, M. D. Hook and L. Balents,
\prb~{\bf 84}, 235126 (2011)

\bibitem{Mullen2015}
K. Mullen, B. Uchoa and D. T. Glatzhofer,
\prl~{\bf 115}, 026403 (2015).

\bibitem{Bian2016}
G. Bian, T.-R. Chang, H. Zheng, S. Velury, S.-Y. Xu, T.
Neupert, C.-K. Chiu, S.-M. Huang, D. S. Sanchez, I. Belopolski, N. Alidoust, P.-J. Chen, G. Chang, A. Bansil,
H.-T. Jeng, H. Lin and M. Z. Hasan,
\prb~{\bf 93}, 121113 (2016).

\bibitem{Yu2015}
R. Yu, H. Weng, Z. Fang, X. Dai, and X. Hu,
\prl~{\bf 115}, 036807 (2015).

\bibitem{Chen2015}
Y. Chen, Y. Xie, S. A. Yang, H. Pan, F. Zhang, M. L.
Cohen and S. Zhang,
Nano. Lett. {\bf 15}, 6974 (2015).

\bibitem{Li2018}
L.~Li, C.~H. Lee, and J.~Gong,
\prl~{\bf 121}, 036401 (2018).

\bibitem{Xu2011}
G. Xu, H. Weng, Z. Wang, X. Dai and Z. Fang,
\prl~{\bf 107}, 186806 (2011).

\bibitem{Nielsen1981}
H. B. Nielsen and M. Ninomiya,
Phys. Lett. B, ~{\bf 105}, 219, (1981).

\bibitem{Hosur2012}
P. Hosur,
\prb~{\bf 86}, 195102 (2012).

\bibitem{Potter2014}
A. C. Potter, I. Kimchi and A. Vishwanath,
Nature Communications ~{\bf 5}, 5161 (2014).

\bibitem{Adler1969}
S. L. Adler,
Phys. Rev. {\bf 177}, 2426 (1969).

\bibitem{Bell1969}
J. S. Bell and R. Jackiw,
Nuovo Cimento A {\bf 60}, 47 (1969).

\bibitem{Zyuzin2012}
A. A. Zyuzin and A. A. Burkov,
\prb~{\bf 86}, 115133 (2012).

\bibitem{Liu2013a}
C.-X. Liu, P. Ye, and X.-L. Qi,
\prb~{\bf 87}, 235306 (2013).



\bibitem{Burkov2014}
A. A. Burkov,
\prl~{\bf 113}, 247203 (2014).

\bibitem{Nielsen1983}
H. B. Nielsen and M. Ninomiya,
Phys. Lett. B, ~{\bf 130}, 389 (1983).

\bibitem{Burkov2011a}
A. Burkov, and L. Balents,
\prl~{\bf 107}, 127205 (2011).

\bibitem{Kitagawa2010}
T. Kitagawa, E. Berg, M. Rudner, and E. Demler,
\prb~{\bf 82}, 235114 (2010).

\bibitem{Lindner2011}
N. H. Lindner, G. Refael, and V. Galitski,
Nat. Phys. {\bf 7}, 490 (2011).

\bibitem{DerekPRL2012}
D. Y. H. Ho and J. Gong, Phys. Rev. Lett. {\bf 109}, 010601 (2012).

\bibitem{Rechtsman2013}
M. C. Rechtsman, J. M. Zeuner, Y. Plotnik, Y. Lumer, D.
Podolsky, F. Dreisow, S. Nolte, M. Segev, and A. Szameit,
Nature (London) {\bf 496}, 196 (2013).

\bibitem{Wang2013Exp}
Y. H. Wang, H. Steinberg, P. Jarillo-Herrero, and N. Gedik, Science, {\bf 342} (6157), (2013).

\bibitem{Rudner2013}
M. S. Rudner, N. H. Lindner, E. Berg and M. Levin,
Phy. Rev. X {\bf 3}, 031005 (2013).

\bibitem{Asboth2014}
J. K. Asboth, B. Tarasinski and P. Delplace,
\prb ~{\bf 90}, 125143 (2014).

\bibitem{Lababidi2014}
M. Lababidi, I. I. Satija and E. Zhao,
\prl ~{\bf 112}, 026805 (2014).


\bibitem{Fulga2016}
I. C. Fulga and M. Maksymenko,
\prb ~{\bf 93}, 075405 (2016).


\bibitem{Zhou2018}
L. Zhou and J. Gong,
\prb ~{\bf 97}, 245430, (2018).


\bibitem{Gong2016}
T. S. Xiong, J. Gong, and J. H. An,  
\prb ~{\bf 93}, 184306 (2016).

\bibitem{Umer2020}
M. Umer, R. W. Bomantara and J. Gong,
\prb ~{\bf 101}, 235438 (2020).


\bibitem{Mciver2020Exp}
J. W. McIver, B. Schulte, F.U. Stein, T. Matsuyama, G. Jotzu, G. Meier and A. Cavalleri, Nat. Phys. {\bf 16}, 38 (2020).

\bibitem{Coldatom}
K. Wintersperger, C. Braun, F.N. \''{U}nal, A. Eckardt,
M.D. Liberto, N. Goldman, I. Bloch, and M. Aidelsburger,
Nat. Phys. 10.1038/s41567-020-0949-y (2020).

\bibitem{Jiang2011}
L. Jiang, T. Kitagawa, J. Alicea, A. R. Akhmerov, D.
Pekker, G. Refael, J. I. Cirac, E. Demler, M. D. Lukin,
and P. Zoller,
\prl~{\bf 106}, 220402 (2011).


\bibitem{Tong2013}
Q.-J. Tong, J.-H. An, J. Gong, H.-G. Luo, and C. H. Oh,
\prb~{\bf 87}, 201109(R) (2013).

\bibitem{RadityaPRL08} 
R. W. Bomantara and J. Gong, Phys. Rev. Lett. {\bf 120}, 230405 (2018).

\bibitem{RadityaPRB08}
R. W. Bomantara and J. Gong, Phys. Rev. B {\bf 98}, 165421 (2018).

\bibitem{Liu2013}
D. E. Liu, A. Levchenko, and H. U. Baranger,
\prl~{\bf 111}, 047002 (2013).

\bibitem{Bomantara2016}
R. W. Bomantara, G. N. Raghava, L. Zhou and J. Gong,
\pre ~{\bf 93}, 022209 (2016).

\bibitem{Bomantara2016a}
R. W. Bomantara and J. Gong,
\prb ~{\bf 94}, 235447 (2016).

\bibitem{Wang2016}
H. Wang, L. Zhou, and Y. D. Chong,
\prb ~{\bf 93}, 144114 (2016).

\bibitem{Gong2016b} L. Zhou, C. Chen, and J. Gong, 
 \prb{\bf 94}, 075443 (2016).

\bibitem{Wang2017}
H.-Q. Wang, M. N. Chen, R. W. Bomantara, J. Gong, and D. Y. Xing,
\prb ~{\bf 95}, 075136 (2017).

\bibitem{Bucciantini2017}
L. Bucciantini, S. Roy, S. Kitamura and T. Oka,
\prb ~{\bf 96}, 041126(R) (2017).

\bibitem{Peri2018}
V. Peri and S. D. Huber,
arXiv:1812.06994v1

\bibitem{Zhu2020}
Y. Zhu, T. Qin, X. Yang, G. Xianlong, and Z. Liang,
Phys. Rev. Research {\bf 2}, 033045 (2020).

\bibitem{NC2017}
Hannes Hübener, Michael A. Sentef, Umberto De Giovannini, Alexander F. Kemper and Angel Rubio, Nature Communications {\bf 8}, 13940 (2017). 

\bibitem{Nathan2015}
F. Nathan and M. S. Rudner,
New J. Phys.~{\bf 17}, 125014 (2015).

\bibitem{Yao2017}
S. Yao, Z. Yan and Z. Wang,
\prb~{\bf 96}, 195303 (2017).

\bibitem{Shirley1965}
J. H. Shirley, Phys. Rev. {\bf 138}, B979 (1965).

\bibitem{Sambe1973}
H. Sambe, \pra ~{\bf 7}, 2203 (1973).

\bibitem{Hockendorf2017}
B. Höckendorf, A. Alvermann and H. Fehske,
J. Phys. A: Math. Theor. {\bf 50}, 295301 (2017).

\bibitem{Leboeuf1990KHM}
P. Leboeuf, J. Kurchan, M. Feingold, and D. P. Arovas, Phys. Rev. Lett. {\bf 65}, 3076 (1990).

\bibitem{Wang2013KHM}
H. Wang, D. Y. H. Ho, W. Lawton, J. Wang, and J. B. Gong, Phys. Rev. E {\bf 88}, 052920 (2013).

\bibitem{Derek2014KHM}
D. Y. H. Ho and J. B. Gong, Phys. Rev. B {\bf 90}, 195419 (2014).

\bibitem{Gallagher2014}
P. Gallagher, M. Lee, J. R. Williams and D. G.-Gordon,
Nature Physics {\bf 10}, 748–752 (2014).

\bibitem{Sumrule}
A. Kundu and B. Seradjeh,
\prl~{\bf 111}, 136402 (2013).

\bibitem{Yap2017}
H. H. Yap, L. Zhou, J.-S. Wang, and J. Gong,
\prb~{\bf 96}, 165443 (2017).

\bibitem{Yap2018}
H. H. Yap, L. Zhou, C. H. Lee, and J. Gong,
\prb~{\bf 97}, 165142 (2018).

\bibitem{Sun2018}
X.-Q.~Sun, M.~Xiao, Tom\'{a}s Bzdusek, S.-C.~Zhang, and S.~Fan,
\prl~{\bf 121}, 196401 (2018).

\end{thebibliography}

\end{document}